\documentclass[12pt]{article} 
\usepackage[sectionbib]{natbib}
\usepackage{array,epsfig,fancyheadings,rotating}
\usepackage[]{hyperref}  
\usepackage{sectsty, secdot}
\sectionfont{\fontsize{12}{14pt plus.8pt minus .6pt}\selectfont}
\renewcommand{\theequation}{\thesection\arabic{equation}}
\subsectionfont{\fontsize{12}{14pt plus.8pt minus .6pt}\selectfont}

\usepackage{amsmath}
\usepackage{amssymb}
\usepackage{amsfonts}
\usepackage{multirow}
\usepackage{amsthm}

\usepackage{mathtools}
\usepackage{booktabs}
\usepackage{enumitem}
\usepackage{array}
\usepackage{bbm}

\usepackage{times}
\usepackage{bm}
\usepackage{multirow}
\usepackage{graphicx}
\usepackage{caption}
\usepackage[titletoc]{appendix}
\usepackage[plain,noend]{algorithm2e}
\usepackage{url} 
\newtheorem{theorem}{Theorem}

\theoremstyle{definition}

\begin{document}


\renewcommand{\baselinestretch}{2}

\renewcommand{\thefootnote}{}
$\ $\par


\fontsize{12}{14pt plus.8pt minus .6pt}\selectfont \vspace{0.8pc}
\centerline{\large\bf Calibration of Inexact Computer Models with Heteroscedastic Errors}
\vspace{.4cm} \centerline{Chih-Li Sung$^1$, Beau David Barber$^2$, Berkley J. Walker$^1$} \vspace{.4cm} \centerline{\it
$^1$Michigan State University, $^2$University of Illinois at Urbana-Champaign}
\vspace{.55cm} \fontsize{9}{11.5pt plus.8pt minus
.6pt}\selectfont

\begin{quotation}
\noindent {\it Abstract:}
Computer models are commonly used to represent a wide range of real systems, but they often involve some unknown parameters. Estimating the parameters by collecting physical data becomes essential in many scientific fields, ranging from engineering to biology. However, most of the existing methods are developed under the assumption that the physical data contains homoscedastic measurement errors. Motivated by an experiment of plant relative growth rates where replicates are available, we propose a new calibration method for inexact computer models with heteroscedastic measurement errors. Asymptotic properties of the parameter estimators are derived, and a goodness-of-fit test is developed to detect the presence of heteroscedasticity.  
Numerical examples and empirical studies demonstrate that the proposed method not only yields accurate parameter estimation, but it also provides accurate predictions for physical data in the presence of both heteroscedasticity and model misspecification. 

\vspace{9pt}
\noindent {\it Key words and phrases:}
Gaussian process, input-dependent noise, plant biology, replication, uncertainty quantification.
\end{quotation}\par

\def\thefigure{\arabic{figure}}
\def\thetable{\arabic{table}}

\renewcommand{\theequation}{\thesection.\arabic{equation}}

\fontsize{12}{14pt plus.8pt minus .6pt}\selectfont

\setcounter{section}{0} 
\setcounter{equation}{0} 

\lhead[\footnotesize\thepage\fancyplain{}\leftmark]{}\rhead[]{\fancyplain{}\rightmark\footnotesize\thepage}

\section{Introduction}
Computer models, which use mathematical representations to simulate real systems, have been widely adopted to understand a real-world feature, phenomenon or event. The applications of computer models range from economics to the physical and biological sciences. For instance, high-fidelity computer simulations are conducted in \cite{mak2017efficient} to study turbulent flows in a swirl injector, which are used in a wide variety of engineering applications such as the design of contemporary liquid rocket engines. A computer model often contains some unknown parameters that represent certain inherent attributes of the underlying systems but cannot be directly controlled or measurable in its physical experiment, which are called \textit{calibration parameters} in the literature \citep{santner2003design}. When its physical experiment is available, these parameters are used to \textit{calibrate} the computer model such that the model simulations agree with characteristics observed in the physical experiment. This process is called \textit{calibration}, and it is of great importance for computer modelers because it not only improves the model prediction, but the estimated value of the calibration parameters also provides some scientific insight which can help modelers better understand the system. For example, the parameters in the cell adhesion study of \cite{sung2018calibration} include kinetic rates and their estimated values provide the information of molecular interactions in the biological system. 

This paper is motivated by the preponderance of computer models used to interpret biological data in plant biology.
For example, the underlying biochemistry facilitating carbon fixation by plants can be determined by calibrating a computer model with rates of photosynthesis measured as a function of light intensity or carbon dioxide concentration 
\citep{sharkey2007fitting,von2013steady}. Computer models are also used to quantify plant metabolic fluxes \citep{ma2014isotopically}. 
Oftentimes gathering and interpreting data from plant science experiments face the same challenges when used in calibration approaches: (1) models are inexact or imperfect due to simplifications or incomplete understanding of the system, and (2) data contain heteroscedastic variance due to limitations in measurement approaches and variability in plant development. A real problem in plant biology involving inexact models and heteroscedastic errors will be illustrated in Section \ref{sec:motivatedexample}. Apart from plant biology, many problems in astronomy also face these two challenges. See, for example, \cite{long2017note} where a sinusoidal model is used to estimate the period of a single periodic variable star but cannot perfectly represent the light curve shape. In these problems, weighted least-squares (WLS) estimators are typically used to estimate the calibration parameters. The WLS estimators, however, can be shown to be generally inconsistent when the computer model is inexact and the error is input-dependent. The details will be given in Section \ref{sec:model}.

In this paper, \textit{we aim to develop a new calibration framework for inexact computer models with replicated experiments potentially having input-dependent errors}, in which we will (i) study how to estimate calibration parameters and make predictions with model uncertainty in the face of heteroscedasticity; (ii) study the asymptotic properties of the estimators; (iii) develop a goodness-of-fit test to detect the presence of heteroscedasticity. Although there has been much work on calibration problems for inexact computer models in the statistics literature (e.g., \cite{kennedy2001bayesian,tuo2015efficient,plumlee2017bayesian,higdon2008computer}), these methods are developed under a homoscedastic assumption,  
which  may in turn  lead to faulty inferences in the presence of heteroscedasticity. On the other hand, recent study by \cite{long2017note} proposes an adaptive estimator which accounts for heteroscedasticity and has lower asymptotic variance than ordinary least-square  and WLS estimators. This method, however, is limited to a linear computer model and requires the assumption that the variances are independent of input variables. Some techniques for addressing heteroscedasticity, such as \cite{binois2018practical} and \cite{ankenman2010stochastic}, could be used for the calibration problem, but the asymptotic properties of estimators have not been systematically studied.

The remainder of this paper is organized as follows. To motivate the following developments, a real problem arising from plant biology is first illustrated in Section \ref{sec:motivatedexample} to establish a more complete background of the problems addressed herein. In Section \ref{sec:model}, a heteroscedastic model is introduced, and the estimation procedure for the calibration parameters and the hyperparameters in the model is developed. Asymptotic properties of the parameter estimators and goodness-of-fit of the heteroscedastic model are presented in Section \ref{sec:inference}. Synthetic examples are illustrated in Section \ref{sec:illustrative}. The proposed framework is applied to the case study of plant relative growth rates in Section \ref{sec:casestudy}. Concluding remarks are given in Section \ref{sec:discussion}. Estimation details, mathematical proofs, supporting figures, an \texttt{R} \citep{R2018} package, and the \texttt{R} \citep{R2018} code for implementation are provided in Supplementary Materials.

\subsection{Illustrative Example}\label{sec:motivatedexample}
Plant relative growth rate \citep{blackman1919compound,hunt1982plantb,hunt1982planta} plays an important role to study the performance of plant productivity as related to environmental stress and disturbance regimes. To calculate the relative growth rate of plants, a relatively simple, yet mechanistically accurate, computer model is commonly used. This model expresses plant biomass after $x$ days, $S(x)$, in the form of an equation:
\begin{equation}\label{eq:computermodel}
    \frac{d S(x)}{dx}=\theta S(x),
\end{equation}
where $\theta$ is a constant and defined as the relative growth rate. This differential equation has the solution $S(x)=S(0)\exp(\theta x)$. In this study the initial plant biomass is set $S(0)=1$. 
The experimental data are observations from plant growth experiments, where the plant biomass is quantified by imaging the photosynthetically active areas of the tissues using a camera system that measured fluorescence emitted from photosynthetically active tissues \citep{murchie2013chlorophyll}. 
To account for technical and biological variation, multiple replicates are conducted in the experiment. 

\begin{figure}[!ht]
    \centering
    \includegraphics[width=0.7\textwidth]{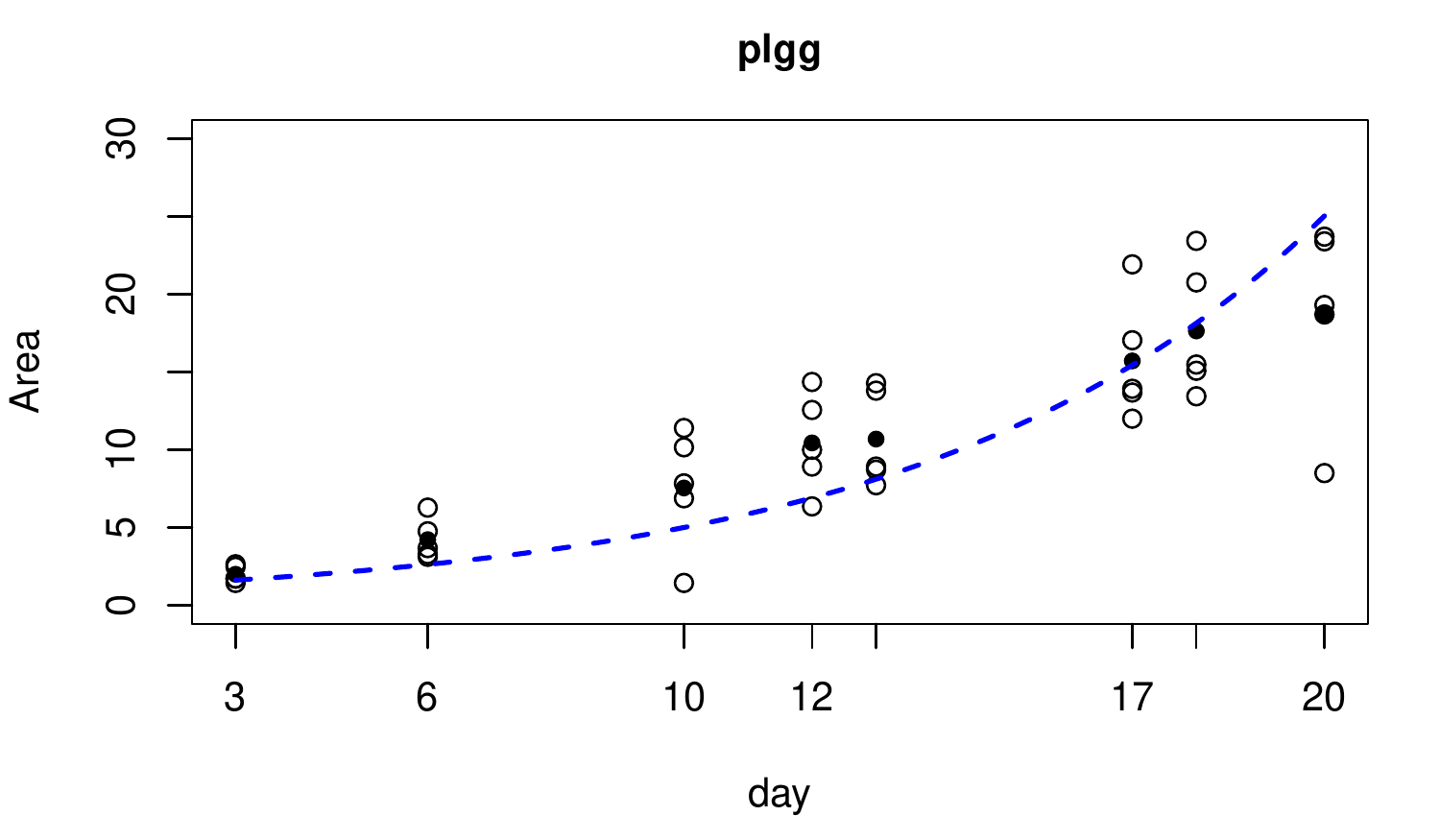}
    \caption{Experimental observations of the plant group \textit{plgg} under high CO$_2$, where the open circles are  replicates with the averaged observation in filled circles at each input location, and the blue dashed line is the computer model with the weighted least-squares estimate.}
    \label{fig:demonstration}
\end{figure}

Figure \ref{fig:demonstration} shows the the photosynthetically active areas of the plant group \textit{plgg} (more details of which will be given in Section \ref{sec:casestudy}) over the three week period, along with the WLS estimates as the blue dashed line. First, it can be seen that the variations of the replicates appear to be different across these three weeks. In particular, the variances appear to be larger after day 6. Secondly, the computer model appears to be misspecified. In particular, we would expect that the model output is close to the averaged observation if the model is correct, but this model seems to underestimate the growth in the early stage and overestimate in the latter stage. This model misspecification could result from simplifications or misunderstanding of the system. Therefore, statistical correction of model predictions is in need which will be developed in the following sections.

\section{Heteroscedastic Modeling for Calibration Problems under Replication}\label{sec:model}
Suppose that $N$ observations are collected from the physical experiments, denoted by $y_1,\ldots,y_N$, and their corresponding inputs are $x_1,\ldots,x_N$, where $x_i\in\chi\subseteq\mathbb{R}^d$. In the case of replication, we further denote $\bar{x}_i$, $i=1,\ldots,n$ as the $n$ unique input locations, where $n<N$, and $y^{(j)}_i$ as the $j$-th output out of $a_i\geq 1$ replicates at the unique location $\bar{x}_i$, and denote its sample mean, $\sum^{a_i}_{j=1}y^{(j)}_i/a_i$, as $\bar{y}_i$. Furthermore, denote $f(x,\theta)$ as the computer model which is a function of the input $x\in\chi\subseteq\mathbb{R}^d$ and the calibration parameter $\theta\in\Theta$ where $\Theta$ is a compact subset of $\mathbb{R}^q$. Here we focus on a deterministic computer model and assume the computer model is known or cheap to evaluate at any input. The generalization to expensive computer models will be discussed in Section \ref{sec:ogp}. Then, the calibration problem for inexact computer models with heteroscedastic errors can be represented as follows,
\begin{equation}\label{eq:model}
    y(x_i)=\zeta(x_i)+\epsilon_i,\quad i=1,\ldots,N,
\end{equation}
where 
\begin{equation}\label{eq:trueprocess}
    \zeta(x_i)=f(x_i,\theta)+b_{\theta}(x_i),
\end{equation}
and $\epsilon_i$ is the measurement or some stochastic error from the real system and independently and identically follows an normal distribution with zero mean and variance $\mathbb{V}[\epsilon_i]=r(x_i)$. The function $\zeta(\cdot)$ is the \textit{true process} of the real system, and the function $b_{\theta}(\cdot)$ in \eqref{eq:trueprocess} is the \textit{discrepancy} (or \textit{bias}) between the true process and the computer model. The inclusion of the discrepancy term in the model is necessary because the computer models are often considered \textit{inexact} or \textit{imperfect}, meaning that even with an optimal calibration parameter, the computer model does not perfectly match the true process, which is referred to as \textit{model uncertainty} in the literature \citep{kennedy2001bayesian}.  Note that the dependence of $b_{\theta}(\cdot)$ on $\theta$ is often suppressed in the literature, but it is included here for clarity. It is also worth noting that 
when $r(x_i)$ is assumed to be constant, this model is a special case of \cite{kennedy2001bayesian}. When $f(x,\theta)$ is a constant mean or a linear function and $b_{\theta}(\cdot)$ follows a Gaussian process model that is independent of $\theta$, the heteroscedastic model is closely related to the models of \cite{ankenman2010stochastic} and \cite{binois2018practical}, where their primary objective is to emulate stochastic simulations, whereas our focus here is on estimation and inference of the calibration parameters as well as statistical correction of model predictions.


The inexact computer models were first discussed by \cite{kennedy2001bayesian} which model the model discrepancy as a Gaussian process (GP) model, and this method has been widely used in many applications (e.g., \cite{higdon2004combining,higdon2008computer,wang2009bayesian,han2009simultaneous}).  
Following the idea of \cite{kennedy2001bayesian}, we assume that the distribution of $b_{\theta}(\cdot)$ is represented by a GP with zero mean and a positive-definite covariance function $c$, so that $b_{\theta}(x_1),\ldots,b_{\theta}(x_N)$ is a multivariate normal distribution with zero mean and covariance matrix $(c(x_i,x_j))_{1\leq i\leq j\leq N}$. A scale $\nu>0$ is commonly separated from a kernel function, $c(x_i,x_j)=\nu k(x_i,x_j;\boldsymbol{\varphi})$, where $\boldsymbol{\varphi}$ are hyperparameters of the kernel. The dependency of $\boldsymbol{\varphi}$ will be suppressed in the rest of the paper for notational simplicity. Typical choices of the kernel function are Gaussian or Mat{\'e}rn kernels which are independent of $\theta$. However, recent studies (e.g., \cite{gramacy2015calibrating,tuo2015efficient,tuo2016theoretical,plumlee2017bayesian}) indicate that these choices of kernel functions may lead to unreasonable calibration parameter estimation due to \textit{unidentifiability} of the calibration parameters. Therefore, in this paper, we consider an \textit{orthogonal kernel function} \citep{plumlee2017bayesian} to avoid the identifiability issue, which will be introduced in Section \ref{sec:ogp}.

Thus, given the noise function $r(x)$, the observations $\mathbf{Y}_N=(y_1,\ldots,y_N)$ follow a multivariate normal distribution,
\begin{equation*}
    \mathbf{Y}_N\sim\mathcal{N}(\mathbf{f}(\theta),\nu(\mathbf{K}_N+\boldsymbol{\Lambda}_N)),
\end{equation*}
where $\mathbf{f}(\theta)=(f(x_1,\theta),\ldots,f(x_N,\theta))^T$, $\mathbf{K}_N$ is an $N\times N$ matrix with $ij$ elements $k(x_i,x_j)$, and $\boldsymbol{\Lambda}_N$ is an $N\times N$ diagonal matrix with diagonal elements $\lambda_1,\ldots,\lambda_N$, where $\lambda_i=r(x_i)/\nu$. Based on the properties of conditional multivariate normal distributions, the predictive distribution of $y(x)$ at a new input setting $x$, $y(x)|\mathbf{Y}_N$, is a normal distribution, $\mathcal{N}(\mu(x),\sigma^2(x))$, where 
\begin{equation*}
    \mu(x)=f(x,\theta)+\mathbf{k}(x)^T(\mathbf{K}_N+\boldsymbol{\Lambda}_N)^{-1}(\mathbf{Y}_N-\mathbf{f}(\theta))
\end{equation*}
and 
\begin{equation}\label{eq:preditivevar}
    \sigma^2(x)=\nu k(x,x) + r(x) - \nu\mathbf{k}(x)^T(\mathbf{K}_N+\boldsymbol{\Lambda}_N)^{-1}\mathbf{k}(x),
\end{equation}
where $\mathbf{k}(x)=(k(x,x_1),\ldots,k(x,x_N))^T$.

The main difference of the model from the others in the literature lies in the the heteroscedastic error $\epsilon_i$ whose variance $r(x)$ is non-constant, while typical homoscedastic cases consider a constant variance, $r(x_i)=\tau^2$. This assumption makes the calibration problem more challenging, because the noise function $r(x)$ is unknown which needs to be estimated. A straightforward and sensible estimate is the sample variance of the replicates at each unique location, that is, $\hat{r}(\bar{x}_i)=\sum^{a_i}_{j=1}(y^{(j)}_{i}-\bar{y}_i)^2/(a_i-1)$. For instance, WLS estimation method is commonly used in practice for calibration problems with heteroscedastic errors, which minimizes the weighted least-squares,
\begin{equation}\label{eq:wls}
    \hat{\theta}_{\rm{WLS}}=\arg\min_{\theta\in\Theta}\sum^n_{i=1}\frac{(\bar{y}_i-f(\bar{x}_i,\theta))^2}{r(\bar{x}_i)},
\end{equation}
where $r(\bar{x}_i)$ is typically estimated by $\hat{r}(\bar{x}_i)$. Practical examples include \cite{antoniewicz2006determination} and \cite{ma2014isotopically} which estimate the calibration parameters $\theta$ (or metabolic flux in their context) by minimizing the weighted least-squares. The estimate $\hat{r}(\bar{x}_i)$ was also used in \cite{ankenman2010stochastic} where they fit a GP model on  the pairs $(\bar{x}_i,\hat{r}(\bar{x}_i))$. This estimate, however, often requires a minimal number of replicates. For example, \cite{ankenman2010stochastic} recommends $a_i\geq10$ replicates for fitting a stochastic GP while \cite{wang2018controlling} recommends $a_i\geq 5$. This is impractical in many applications because physical data from a real system is often time-consuming or too costly to collect. In addition, the predictive variance \eqref{eq:preditivevar} still requires the value of $r(x)$ at the new input setting $x$ but physical data at this input setting is not directly available for estimation. To this end, we employ the latent log-variance GP proposed by \cite{binois2018practical} to model the noise function $r(\cdot)$, which does not require a minimal number of the replicates and is computationally efficient under replication. This model will be briefly reviewed in the next subsection.

As a matter of fact, even when $r(\bar{x}_i)$ is known, it can be shown that the WLS estimator $\hat{\theta}_{\rm{WLS}}$ is generally inconsistent when the computer model is inexact. A theorem that shows the result is given below, and its proof is provided in Supplementary Material \ref{append:theorem_inconsistency}. Suppose $x_i\sim F_X$ independent across $i$. We first define the true parameter as
\begin{equation}\label{eq:truetheta_Long}
    \theta^*=\arg\min_{\theta\in\Theta}\mathbb{E}[(\zeta(X)-f(X,\theta))^2],
\end{equation}
which is the parameter of the best fitting least squares. The definition is a simple extension of \cite{long2017note} where he considered $f$ as a linear function so $\theta^*$ is the slope of the best fitting least squares line. When $X$ is uniformly distributed, the definition is equivalent to the $L_2$-calibration parameter in \cite{tuo2015efficient}, \cite{tuo2016theoretical}, and \cite{wong2017frequentist}, where they define the true parameter as the $L_2$ distance projection of $\theta$. That is,
\begin{equation}\label{eq:truetheta}
    \theta_{L_2}^*=\arg\min_{\theta\in\Theta}\|\zeta(\cdot)-f(\cdot,\theta)\|^2_{L_2(\chi)},
\end{equation}
where $\|g\|_{L_2(\chi)}=\left(\int_{\chi}g(x)^2{\rm{d}}x\right)^{1/2}$.

\begin{theorem}\label{thm:wls_consistency}
Assume $0<r(x)<\infty$ for any $x\in\chi$. Let $$\theta'=\arg\min_{\theta\in\Theta}\mathbb{E}\left[\frac{(\zeta(X)-f(X,\theta))^2}{r(X)}\right].$$ and assume it has a unique solution. Then, $\hat{\theta}_{\rm{WLS}}$ converges almost surely to $\theta'$. 
\end{theorem}

The theorem implies that when $r(x)$ is non-constant, or equivalently, $r(x)$ is input-dependent, $\hat{\theta}_{\rm{WLS}}$ is generally inconsistent with the true parameter $\theta^*$. 

\subsection{Latent Variable Process for Modeling $r(\cdot)$}
Since $r(\cdot)=\nu\lambda(\cdot)$, we instead model $\lambda(\cdot)$ and $r(\cdot)$ can then be obtained by multiplying the scale $\nu$. Denote $\boldsymbol{\Lambda}_n=(\lambda(\bar{x}_1),\ldots,\lambda(\bar{x}_n))$ and $\mathbf{A}_n=\text{diag}(a_1,\ldots,a_n)$. Similar to \cite{goldberg1998regression} which suggests a GP prior for $\log \lambda(\cdot)$, \cite{binois2018practical} models $\log\lambda_1,\ldots,\log\lambda_n$ as derived quantities obtained via the predictive mean of a regularizing GP on new latent variables, $\delta_1,\ldots,\delta_n$: 
\begin{equation*}
    \log\boldsymbol{\Lambda}_n=\mathbf{K}_{(g)}\left(\mathbf{K}_{(g)}+g\mathbf{A}^{-1}\right)^{-1}\boldsymbol{\Delta}_n,
\end{equation*}
where $\boldsymbol{\Delta}_n=\text{diag}(\delta_1,\ldots,\delta_n)$, $\mathbf{K}_{(g)}=\left(k_{(g)}(\bar{x}_i,\bar{x}_j)\right)_{1\leq i,j\leq n}$ is the kernel matrix whose nugget is $g$, and $\boldsymbol{\Delta}_n\sim\mathcal{N}(0,\nu_{(g)}(\mathbf{K}_{(g)}+g\mathbf{A}_n^{-1}))$. $k_{(g)}$ is a kernel function of this noise process, and it contains some hyperparameters which are denoted by $\boldsymbol{\phi}$. Typical kernels such as Gaussian or Mat{\'e}rn kernels can be used here. The latent variable $\boldsymbol{\Delta}_n$ are unknown and treated as additional parameters, which will be estimated in Section \ref{sec:estimation} along with $\boldsymbol{\phi}$ and nugget $g$. The predictive value of $\log \lambda(x)$ at an new input $x$ can then be obtained by $\log \lambda(x)=\mathbf{k}_{(g)}(x)^T\left(\mathbf{K}_{(g)}+g\mathbf{A}_n^{-1}\right)^{-1}\boldsymbol{\Delta}_n$, where $\mathbf{k}_{(g)}(x)=(k_{(g)}(x,\bar{x}_1),\ldots,k_{(g)}(x,\bar{x}_n))^T$. We refer more details of the latent log-variance GP to \cite{binois2018practical}.

\subsection{Orthogonal Gaussian Process for Modeling $b_{\theta}(\cdot)$}\label{sec:ogp}
Although the GP modeling of \cite{kennedy2001bayesian} (referred to as KO in the rest of the paper) for $b_{\theta}(\cdot)$ has been widely used, recent studies have raised concerns about its identifiability issue of the calibration parameters \citep{loeppky2006computer,bayarri2007computer,bayarri2007framework,han2009simultaneous,gramacy2015calibrating,tuo2015efficient,tuo2016theoretical,plumlee2017bayesian,wong2017frequentist}. In particular, \cite{tuo2016theoretical} point out that the KO estimator is asymptotically inconsistent when the true parameter is defined as \eqref{eq:truetheta}. \cite{plumlee2017bayesian} further points out that the GP modeling of the bias $b_{\theta}(\cdot)$ should depend on $\theta$, but the one in KO does not. Therefore, \cite{plumlee2017bayesian} accounts for the definition of \eqref{eq:truetheta_Long} and provides an alternative GP modeling by \textit{orthogonalizing} the model bias to avoid mixing the GP and the definition of the parameter. The idea is to create an alternative kernel function of the GP based on the orthogonality condition,
$$
\int_{\chi}\frac{\partial }{\partial\theta}f(\xi,\theta)b_{\theta}(\xi){\rm{d}} F_X(\xi),
$$
which he shows is a necessary condition to minimize the least squares in \eqref{eq:truetheta_Long}. We briefly introduce the orthogonal GP as follows.

Suppose that $k_0(\cdot,\cdot)$ is any valid kernel function on $\chi\times\chi$ and is independent of $\theta$, such as Gaussian or Mat{\'e}rn kernels. Under the definition of \eqref{eq:truetheta_Long}, \cite{plumlee2017bayesian} suggests a GP on the bias with the \textit{orthogonal kernel function}, 
\begin{equation}\label{eq:OGP}
    k(x_i,x_j)=k_0(x_i,x_j)-h_{\theta}(x_i)^TH_{\theta}^{-1}h_{\theta}(x_j),
\end{equation}
where 
\begin{equation*}
    h_{\theta}(x)=\int_{\chi}\frac{\partial }{\partial\theta}f(\xi,\theta)k_0(x,\xi){\rm{d}} F_X(\xi)
\end{equation*}
and 
\begin{equation*}
    H_{\theta}=\int_{\chi}\int_{\chi}\frac{\partial }{\partial\theta}f(\xi_1,\theta)\left(\frac{\partial }{\partial\theta}f(\xi_2,\theta)\right)^Tk_0(\xi_1,\xi_2){\rm{d}} F_X(\xi_1){\rm{d}} F_X(\xi_2).
\end{equation*}
Note that the orthogonal kernel function $k(\cdot,\cdot)$ is dependent of $\theta$ but here it is suppressed for notational simplicity.

In practice, there are two major difficulties to evaluate $k(\cdot,\cdot)$. The first is the integrals in $h_{\theta}$ and $H_{\theta}$ which are often difficult to solve. This can be addressed by using the stochastic average approximation, such as Monte Carlo integration \citep{caflisch1998monte}. For example, one can draw $m$ samples, $\xi_1,\ldots,\xi_m$, from $F_X$, and then approximate $h_{\theta}(x)$ by
\begin{equation*}
    h_{\theta}(x)\approx\frac{1}{m}\sum^m_{i=1}\frac{\partial }{\partial\theta}f(\xi_i,\theta)k_0(x,\xi_i).
\end{equation*}
Second, $k(\cdot,\cdot)$ requires the evaluations of the computer model, $f(x,\theta)$, and its gradient, $\partial f(x,\theta)/\partial\theta$, at any input pair $(x,\theta)\in\chi\times\Theta$, but they are not generally obtainable, because computer simulations can be computationally demanding (e.g., the high-fidelity simulation in \cite{mak2017efficient}). A common approach is to run a computer experiment with various inputs and build an cheaper \textit{emulator} for the actual computer simulations, for which GP modeling is often used \citep{sacks1989design,santner2003design}. Thus, the predictive distribution of the emulator can be taken as a fixed probabilistic definition of $f(\cdot,\cdot)$, and hence the definition of orthogonal kernel function is modified accordingly. We refer more details to \cite{plumlee2017bayesian}.

\subsection{Parameter Estimation}\label{sec:estimation}
The estimation procedure for the model parameters herein is based on maximum likelihood estimation. This procedure is developed along the lines described in \cite{binois2018practical}, which develop computationally efficient inference and prediction for a heteroscedastic GP when replication is present. 
The model parameters include the calibration parameters $\theta$, the hyperparameters of the two GPs, $\boldsymbol{\varphi},\boldsymbol{\phi},g$, and the latent variables $\delta_1,\ldots,\delta_n$. Conditional on the parameters $\theta,\boldsymbol{\varphi},\boldsymbol{\phi},g,\delta_1,\ldots,\delta_n$, the scales $\nu$ and $\nu_{(g)}$ both have plug-in MLEs: $\hat{\nu}=N^{-1}\left(\mathbf{Y}_N-\mathbf{f}(\theta)\right)^T\left(\mathbf{K}_N+\boldsymbol{\Lambda}_N\right)^{-1}\left(\mathbf{Y}_N-\mathbf{f}(\theta)\right)$ and $\hat{\nu}_{(g)}=n^{-1}\boldsymbol{\Delta}_n^T\left(\mathbf{K}_{(g)}+g\mathbf{A}_n^{-1}\right)^{-1}\boldsymbol{\Delta}_n$.
The log-likelihood conditional on $\hat{\nu}$ and $\hat{\nu}_{(g)}$ is then
\begin{align}\label{eq:likelihood}
    \log L=&-\frac{N}{2}\log 2\pi-\frac{N}{2}\log\hat{\nu}-\frac{1}{2}\log|\mathbf{K}_N+\boldsymbol{\Lambda}_N|-\frac{N}{2}\nonumber\\
    &-\frac{n}{2}\log 2\pi-\frac{n}{2}\log\hat{\nu}_{(g)}-\frac{1}{2}\log|\mathbf{K}_{(g)}+g\mathbf{A}_n^{-1}|-\frac{n}{2},
\end{align}
where the top line above is the mean-field component and the bottom line is the variance-field component. While optimizing the log-likelihood can be computationally demanding when $N$ is large, because the inverse and determinant of $\mathbf{K}_N+\boldsymbol{\Lambda}_N$ requires $O(N^3)$ computations, the computation complexity can be efficiently reduced from $O(N^3)$ to $O(n^3+N)$ by using the Woodbury identity \citep{harville1998matrix}, which essentially only depends on the number of the unique input locations.
We leave the details to Supplementary Material \ref{append:gradient}. The parameters can then be efficiently estimated by maximizing the log-likelihood via an optimization algorithm. Following the idea of \cite{binois2018practical}, since the gradient of the log-likelihood is available in a closed form, which is provided in Supplementary Material \ref{append:gradient}, we use a Newton-like optimization method to maximize the log-likelihood. An \texttt{R} package for the estimation procedure is available in the supplementary materials, which is via modifications to the source code of \texttt{R} package \texttt{hetGP} \citep{hetGP2019}. In particular, the optimization is done by the \texttt{optim} library with \texttt{method="L-BFGS-B"}, which performs a quasi-Newton optimization method of \cite{byrd1995limited}. When the gradient, $\partial f(x,\theta)/\partial\theta$, in the orthogonal function of \eqref{eq:OGP} is not available, the function \texttt{gradient} in the \texttt{R} package \texttt{rootSolve} \citep{RrootSovle} is used to approximate the gradient.

\section{Inference and Goodness-of-fit}\label{sec:inference}
In this section, we first study the asymptotic properties for the maximum likelihood estimators obtained in Section \ref{sec:estimation}, which are important for calibration problems as the inference of the calibration parameter is of great interest. Then, a goodness-of-fit statistic is introduced to detect the presence of heteroscedasticity. These theoretical results will be applied to the case study in Section \ref{sec:casestudy}.

\subsection{Asymptotic Properties}\label{sec:asym_prop}
Denote all the model parameters as $\boldsymbol{\omega}=(\theta,\boldsymbol{\psi},\nu,\boldsymbol{\phi},g,\nu_{(g)},\delta_1,\ldots,\delta_n)$ and their estimators as $\hat{\boldsymbol{\omega}
}_N$. Asymptotic results are presented here to show that $\hat{\boldsymbol{\omega}}_N$ is asymptotically normally distributed as $N$ becomes sufficiently large. The regularity conditions and proofs are given in Supplementary Material \ref{append:theorem_asymptotic}. 

\begin{theorem}\label{thm:asymptotic}
Under the regularity conditions in Supplementary Material \ref{append:theorem_asymptotic}, the maximum likelihood estimators $\hat{\boldsymbol{\omega}}_N$ are asymptotically consistent and normal as $N\rightarrow\infty$,
\begin{equation*}
    \mathbf{B}_N(\boldsymbol{\omega})^{1/2}(\hat{\boldsymbol{\omega}}_N-\boldsymbol{\omega})\overset{d}{\longrightarrow}\mathcal{N}(0,I_m),
\end{equation*}
where $I_m$ is the $m\times m$ identity matrix, $m$ is the size of the vector $\boldsymbol{\omega}$, and $\mathbf{B}_N(\boldsymbol{\omega})$ is the information matrix whose closed-form expression is provided in Supplementary Material \ref{append:theorem_asymptotic}.
\end{theorem}

Hence, by the theorem, an approximate $(1-\alpha)\times 100\%$ confidence region of $\theta$ can be constructed as 
$$\left\{\theta\in\Theta\subseteq\mathbb{R}^q|\left(\mathbf{U}\hat{\boldsymbol{\omega}}_N-\theta\right)^T\left(\mathbf{U}\mathbf{B}_N(\hat{\boldsymbol{\omega}}_N)^{-1}\mathbf{U}^T\right)^{-1}\left(\mathbf{U}\hat{\boldsymbol{\omega}}_N-\theta\right)\leq\chi^2_{q,1-\alpha}\right\},
$$
where $\mathbf{U}$ is a $q\times m$ matrix composed of the first through the $q$-th row of the $m$-dimensional identity matrix, and $\chi^2_{q,1-\alpha}$ is the $(1-\alpha)$-quantile of a chi-squared distribution with $q$ degrees of freedom. 

\subsection{Goodness of Fit: Heteroskedasticity Test}\label{sec:het_test}
As the main assumption of the proposed model is the heteroskedastic assumption, it is essential to develop a hypothesis test to detect the presence of heteroskedasticity. There are a variety of test procedures proposed in the literature to detect heteroskedasticity. See, for example, \cite{hildreth1968some,harvey1976estimating,godfrey1978testing,koenker1982robust,newey1987asymmetric}. However, these procedures are developed under the assumption that the specification of the regression function, or the computer model in our context, is correct. These test procedures may falsely indicate the presence of heteroskedasticity if the computer model is incorrect. One exception is the heteroscedastic test of \cite{lee1992heteroskedasticity}, which is robust to the regression function misspecification. This test, however, is limited to detect a linear specification of measurement errors.

Given the model proposed herein, we can provide a more flexible heteroscedastic test for an inexact computer model. In particular, in the proposed model we have $r(x)=\nu\lambda(x)$ and $\log \lambda(x)=\mathbf{k}_{(g)}(x)^T\left(\mathbf{K}_{(g)}+g\mathbf{A}_n^{-1}\right)^{-1}\boldsymbol{\Delta}_n$. Then, under homoscedasticity, the variance function $r(x)$ is constant over $x$, which implies that all of the latent variables, $\delta_1,\ldots,\delta_n$, are equal to zero simultaneously. Therefore, a testable hypothesis to detect heteroskedasticity is 
$$
H_0:\delta_1=\cdots=\delta_n=0\quad\text{v.s.}\quad H_1: \text{ at least one }\delta_i\text{ is non-zero}.
$$
Based on the asymptotic results of Theorem \ref{thm:asymptotic}, a test statistic for the null hypothesis is given in the next theorem. The proof can be done by Slutsky's theorem.

\begin{theorem}
Under the regularity conditions in Theorem \ref{thm:asymptotic}, the heteroscedastic test statistic
$$
\left(\mathbf{H}\hat{\boldsymbol{\omega}}_N\right)^T\left(\mathbf{H}\mathbf{B}_N(\boldsymbol{\omega})^{-1}\mathbf{H}^T\right)^{-1}\left(\mathbf{H}\hat{\boldsymbol{\omega}}_N\right)\overset{d}{\longrightarrow}\chi^2_{n}
$$
for $N$ sufficiently large under the null hypothesis, where $\mathbf{H}$ is an $n\times m$ matrix composed of the last $n$ rows of the $m$-dimensional identity matrix.
\end{theorem}
In a finite sample application, $\mathbf{B}_N(\boldsymbol{\omega})$ can be estimated by $\mathbf{B}_N(\hat{\boldsymbol{\omega}}_N)$, which can be shown to be consistent under the regularity conditions.

\section{Numerical Study}\label{sec:illustrative}
In this section, numerical experiments are conducted to examine the calibration performance of the proposed method, including one computer model with one calibration parameter and one computer model with three calibration parameters. 
In the implementation of the proposed method, Mat{\'e}rn kernels are chosen for $k_{(g)}$, which have the form
\begin{equation*}
    k_{(g)}(x, y) = \left(1+\frac{\sqrt{5}}{\phi} \|x-y\|  +\frac{5}{3\phi^2}\|x-y\|^2\right) \exp\left(-\frac{\sqrt{5}}{\phi}\|x-y\|\right),
\end{equation*}
and the Mat{\'e}rn kernels with hyperparameter $\boldsymbol{\varphi}$ are chosen for $k_0$ in \eqref{eq:OGP} to derive the orthogonal kernel $k$.

\subsection{Example  with One Calibration Parameter}\label{sec:example_1d}
We consider an example adapted from \cite{tuo2015efficient}. Assume that the input $x$ is uniformly distributed on $[0,2\pi]$, the true process is $\zeta(x)=\exp(x/10)\sin x$,
and the observations are given by $y_i=\zeta(x_i)+\epsilon_i$, where $\epsilon_i$ is independently normally distribution with zero mean and the variance $r(x_i)=(0.01+0.2(x_i-\pi)^2)^2$.
Suppose that the computer output is given by the function $f(x,\theta)=\zeta(x)-\sqrt{\theta^2-\theta+1}(\sin\theta x + \cos\theta x)$.
There does not exist a real number $\theta$ such that $f(\cdot,\theta)=\zeta(\cdot)$ because $\sqrt{\theta^2-\theta+1}(\sin\theta x + \cos\theta x)$ is always positive for any $\theta$. Thus, this computer model is \textit{inexact} because even with the optimal setting $\theta^*$, there still exists discrepancy between $f(\cdot,\theta^*)$ and $\zeta(\cdot)$. The true parameter in \eqref{eq:truetheta_Long} can be calculated by minimizing the $L_2$ distance as in \eqref{eq:truetheta} since $x$ is assumed to be uniformly distributed, which gives $\theta^*\approx -0.1789$.

\begin{figure}[b!]
    \centering
    \includegraphics[width=0.9\textwidth]{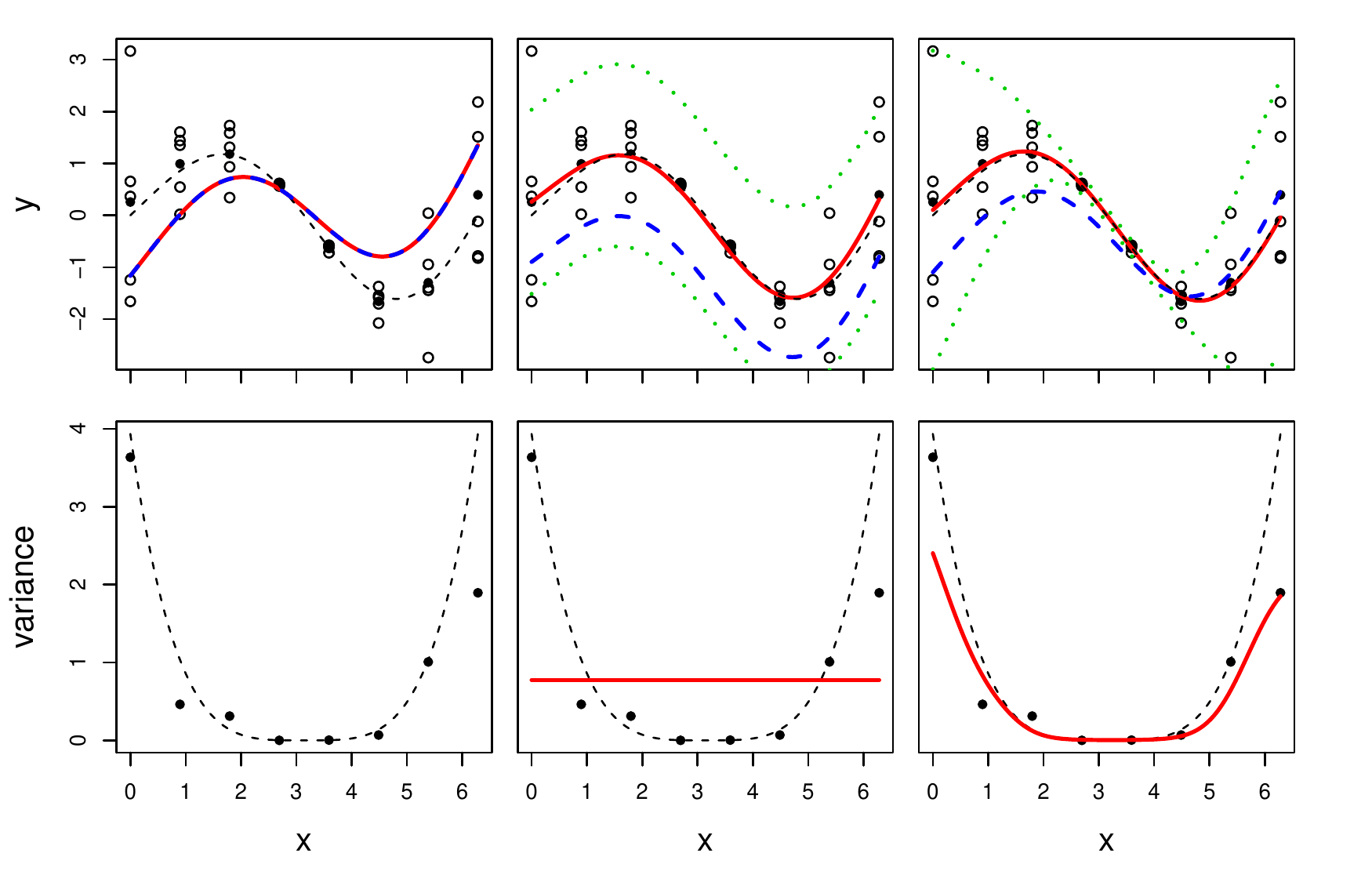}
    \caption{Illustration of three methods: (left) WLS; (middle) KO; (right) the proposed method. Upper panels represent the replicates as open circles with the averaged observation $\bar{y}_i$ in filled circles at each unique input location, the true process as a black dashed line, the computer model $f(\cdot,\hat{\theta})$ as a blue dashed line, and the prediction mean curve as a red solid line, with 95\% prediction intervals in green dotted lines. Lower panels represent the sample variance $\hat{r}({\bar{x}_i})$ as black points, the true variance process as a dashed line, and the fitted variance process as a red solid line.}
    \label{fig:example2demo}
\end{figure}

In this numerical study, eight unique input locations are selected with equal space in $[0,2\pi]$, and 5 replicates are generated at each unique location, that is, $a_1=\ldots=a_8=5$. Figure \ref{fig:example2demo} demonstrates the simulated data, in which three different methods are performed, which are: (left) the WLS estimator; (middle) the homoscedastic modeling, which is the frequentist version of the KO approach; (right) our proposed method. The calibration parameter estimates are -0.2784, 0.2674, and -0.1727, respectively. 
In this example, our proposed method provides a more accurate parameter estimate (the true parameter is $\theta^*\approx -0.1789$), which also can be seen from the upper panels, where the computer model with the estimate is closer to the true process than other two methods in the sense of $L_2$ distances. 
Figure \ref{fig:example2demo} also shows that the WLS yields inaccurate predictions for physical data, and the KO approach suggests unreasonably wide prediction intervals due to the constant variance assumption. On the other hand, the proposed method not only provides a more accurate parameter estimate, but it also provides more accurate predictions as well as more reasonable prediction intervals by recovering the variance process. 

We conduct the simulation 100 times to examine the performance of our proposed method (labeled \textsf{HetOGP}), in comparison with the weighted least-squares (labeled \textsf{WLS}), homoscedastic modeling with a Mat{\'e}rn kernel (labeled \textsf{HomGP}), which is the frequentist version of KO approach, homoscedastic modeling with an orthogonal kernel (labeled \textsf{HomOGP}), which is the frequentist version of the calibration approach in \cite{plumlee2017bayesian}, and heteroscedastic modeling with a Mat{\'e}rn kernel (labeled \textsf{HetGP}), which is close to the model in \cite{binois2018practical}. Three main metrics are used for the comparison: (i) estimation bias, $\hat{\theta}-\theta^*$; (ii) root mean squared errors (RMSEs) based on 101 test equal-spaced locations in $[0,1]$, $\left(\sum^{101}_{i=1}(\zeta(x_i)-\hat{y}(x_i))^2/101\right)^{1/2}$, where $\hat{y}(x)$ is the prediction mean for the input $x$; (iii) predictive score, which is a scoring rule provided by Equation (27) of \cite{gneiting2007strictly} that combines prediction means and variances. Since the true distribution of $y_i$ is known in the simulation setting, the predictive score has the form as follows, 
\begin{equation*}
    -\frac{1}{100}\sum^{100}_{i=1}\left(\frac{\zeta(x_i)-\hat{y}(x_i)}{\hat{\sigma}^2(x_i)}\right)^2-\frac{r(x_i)}{\hat{\sigma}^2(x_i)}-\log\hat{\sigma}^2(x_i),
\end{equation*}
where $\hat{\sigma}^2(x)$ is the prediction variance for the input $x$.

\begin{figure}[h!]
    \centering
    \includegraphics[width=\textwidth]{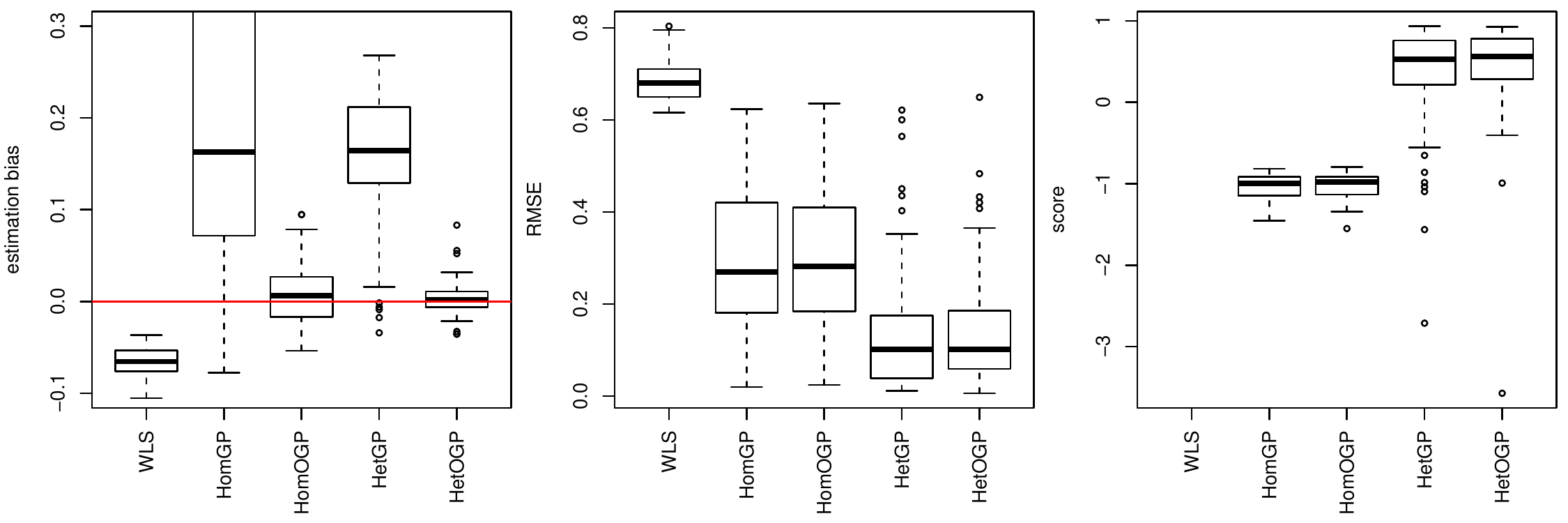}
    \caption{The comparison of estimation and prediction performance. The left panel represents the estimation bias of the calibration parameter, with the red horizontal line indicating zero bias. The middle panel shows their root mean squared errors, and the right panel represents their predictive scores.}
    \label{fig:example2comparison}
\end{figure}

Figure \ref{fig:example2comparison} shows the results for the five methods based on the 100 simulations. The predictive score of \textsf{WLS} is not available because the prediction variances at unobserved input locations are not available for the WLS. First, from the left panel, it can be seen that our proposed method (\textsf{HetOGP}) outperforms the other methods in terms of calibration parameter estimation.
The estimates of \textsf{WLS} are very biased when the computer model is inexact, which is consistent with the result in Theorem \ref{thm:wls_consistency}. \textsf{HomOGP} gives relatively unbiased estimates, but the variation of the estimates is larger than \textsf{HetOGP}. From the middle and right panels, it shows that heteroscedastic modeling-based methods (\textsf{HetGP} and \textsf{HetOGP}) are better than \textsf{WLS} and homoscedastic modeling-based methods (\textsf{HomGP} and \textsf{HomOGP}) in terms of prediction performation. The reason of the poor predictive scores of \textsf{HomGP} and \textsf{HomOGP} is that, as shown in Figure \ref{fig:example2demo}, the homoscedastic modeling yields unreasonable wide prediction intervals when heteroscedasticity is present. \textsf{HetGP} results in superior prediction accuracy and prediction scores, but the estimates are quite off from the true parameter. 
On the other hand, our proposed method (\textsf{HetOGP}), which models the discrepancy function using an orthogonal GP to avoid the identifiability issue, provides more accurate parameter estimates. 
We also construct the 95\% confidence intervals for the calibration parameter based on the result of Theorem \ref{thm:asymptotic}, and 92 out of the 100 simulations cover the true parameter, which is close to the nominal coverage 95\%. In terms of computational cost, all the methods here are implemented within 3 seconds, on a laptop with 2.6 GHz CPU and 16 GB of RAM. Based on the estimation and prediction results, it suggests that our proposed method is more appropriate for the calibration problem in the face of heteroscedasticity and inexact computer models, which provides more accurate parameter estimates along with high prediction accuracy and prediction scores.

\subsection{Example with Three Calibration Parameters}
In this subsection, we consider a calibration problem where two input variables and three calibration parameters are involved in a computer model. This example is adapted from \cite{plumlee2017bayesian}. Assume that the input $x\in\mathbb{R}^2$ is uniformly distributed on $[0,1]^2$, the true process is $\zeta(x)=4x_1+x_1\sin(5x_2)$, and the observations are given by $y_i=\zeta(x_i)+\epsilon_i$, where $\epsilon_i$ is independently normally distribution with zero mean and the variance $r(x_i)=0.01\exp(-10\sin(x_1\pi)\cos(x_2\pi))$.
Suppose that the computer output is given by the function $f(x,\theta)=\theta_1+\theta_2x_1+\theta_3x_2$, where $\theta=(\theta_1,\theta_2,\theta_3)\in\Theta=[0,1]^3$. By minimizing the $L_2$ distance as in \eqref{eq:truetheta}, we have $\theta^*\approx(0.50,4.14,-1.00)$.

Similar to the previous subsection, we conduct the simulation 100 times, where each simulation uses a two-dimensional Latin Hypercube sample (LHS, \cite{mckay1979comparison}) of size 30 on the unit cube for designing the input $x$. In this study we consider three different numbers of replicates, $\{2,5,10\}$, for each unique input setting $x$, leading to $N\in\{60,150,300\}$. The estimation results are summarized in Figure \ref{fig:example3comparison}, which shows the boxplots of estimation bias arranged by numbers of replicates (three groups of five from left to right) for each calibration parameter. The results show that the proposed method (\textsf{HetOGP}) provides more accurate estimates than other four methods for each of the three calibration parameters. \textsf{HomGP} and \textsf{HomOGP} provides relatively unbiased estimates, but the estimate variances are much larger than \textsf{HetOGP}. 
The estimates of the proposed method are more accurate with lower variance by the increase of the replicates, which agrees with the asymptotic result in Theorem \ref{thm:asymptotic}.

\section{Case Study: Estimation of Plant Growth Rate}\label{sec:casestudy}


\subsection{Plant Growth Experiment}
In this experiment, the growth of three groups of plants grown under two different carbon dioxide concentrations was analyzed over a three-week period. The plant species was \textit{Arabidopsis thaliana}, a common experimental plant that is easily manipulated genetically and grows along a flat plane, making growth analysis by overhead imaging of chlorophyll fluorescence possible. 


The three groups of plants differ by the presence or absence of certain genes involved in photorespiration.
The different plant groups tested lacked distinct steps involved in photorespiration, either a critical enzymatic interconversion step (\textit{glyk}) \citep{boldt2005d}, or a transporter (\textit{plgg}) \citep{pick2013} which can be circumvented via other transport mechanisms \citep{walker2016physiological,south2017bile}. These groups were compared to wild type plants (WT), which have a fully functioning photorespiratory pathway. 


\subsection{Calibration Results}
We leverage the statistical developments to investigate the plant relative growth rates for the plants grown under ambient and high CO$_2$ concentrations, with the three plant groups: \textit{glyk}, \textit{plgg}, and WT. The experimental data consists of the total projected areas of the plants at 8 unique time points, as the input variable $x\in[0,20]$, and for each unique time point, three to five replicates are measured. The computer model is as described in \eqref{eq:computermodel}, which shares the same input variable $x$ (time) and has a calibration parameter, the relative growth rate, $\theta\in[0,1]$. 
The calibration results using the experimental data under ambient and high CO$_2$ concentrations are presented in Figures \ref{fig:realcalibration_ambient} and \ref{fig:realcalibration_hi}, respectively. First, it can be seen that in the experimental data under both ambient and high CO$_2$ concentrations, the variances tend to increase as the time increases.
By performing the heteroscedasticity test developed in Section \ref{sec:het_test}, the p-values for the three groups are given in Table \ref{tbl:estimatedparameter}, which shows that all of the p-values are less than 0.05, indicating that  heteroscedastic modeling is essential for this data. 
Our approach takes into account the heteroscedasticity and provides the fitted variance process (as the red lines in the middle panels), which in turn gives sensible prediction intervals (as the green dotted lines in the top panels). Moreover, bottom panels present the fitted discrepancy function, $b_{\hat{\theta}}(x)$, with 95\% pointwise confidence intervals based on the orthogonal GP modeling in Section \ref{sec:ogp}. The discrepancy functions show that the computer model is imperfect as expected, especially for the data under CO$_2$ ambient concentrations (Figure \ref{fig:realcalibration_ambient}), which may suggest that a quadratic polynomial is needed in the computer model. Future studies of plant growth rates may require more complex models if debiasing the computer model is of interest. 
The proposed method not only gives reasonable prediction means and intervals for the experimental data, but it also provides the model discrepancy for computer modelers.

\begin{table}[t!]
    \centering
\begin{tabular}{ c|c|c|c|c }
\toprule
CO$_2$  & \multirow{2}{*}{Group} & \multicolumn{2}{c|}{Relative Growth Rate}  &Het. Test\\  \cline{3-5}
Concentration & & Estimate & 95\% Confidence Interval & p-value\\
\midrule
\multirow{3}{*}{Ambient}  & \textit{glyk} & 0.1590 & [0.1512, 0.1668]&0.0015\\
 & \textit{plgg} & 0.1581 &  [0.1515, 0.1647]& 0.0205\\
 & WT & 0.1780 & [0.1697, 0.1864]& $<$0.0001\\ \midrule
\multirow{3}{*}{High}  & \textit{glyk} & 0.2128 & [0.1947, 0.2308]&$<$0.0001\\
 & \textit{plgg} & 0.2428 & [0.2364, 0.2493] &$<$0.0001\\
 & WT & 0.2295 & [0.2197, 0.2394] &$<$0.0001\\
\bottomrule
\end{tabular}
\caption{Estimated relative growth rates and p-values of heteroscedastic tests.}
\label{tbl:estimatedparameter}
\end{table}


The estimated relative growth rates, $\hat{\theta}$, are reported in Table \ref{tbl:estimatedparameter}, where the confidence intervals are constructed based on the asymptotic normality result in Theorem \ref{thm:asymptotic}. First, we observe that relative growth rates under ambient CO$_2$ concentrations are slower than under high CO$_2$ concentrations across all plant groups. This is expected from the biological perspectives, because CO$_2$/O$_2$ under ambient CO$_2$ concentrations is low enough to drive high rates of photorespiration, which consumes energy and releases previously fixed carbon, decreasing growth. Secondly, the group \textit{glyk} and \textit{plgg} have slower relative growth rates than WT under ambient CO$_2$ concentrations, which is consistent with either a disruption in growth generally or specifically in photorespiration. 
These calibration parameter estimates and confidence intervals provide insight into the values of the relative growth rates of different plant groups, which are difficult to determine by physical experiments due to the limitation of the existing experimental techniques.



\section{Summary and Concluding Remarks}\label{sec:discussion}
Calibration of computer models plays a crucial role in many scientific fields where computer models are essential to predict the reality. The existing methods in the statistics literature, however, mainly focus on calibration with homoscedastic errors. Motivated by an experiment in plant biology, where the noise levels can vary dramatically across different input locations, we introduce a new calibration method to address the heteroscedasticity, where a latent variable process is used to model the error variance and an orthogonal Gaussian process is used to model the misspecification of a computer model. An \texttt{R} package is available for implementing the proposed method. We also study the asymptotic properties of the estimators and provide a goodness-of-fit statistic to detect the presence of heteroscedasticity.  Our numerical studies demonstrate that when the errors are not homoscedastic, our proposed method not only successfully estimates calibration parameters accurately, but it also provides accurate predictions for a real system. The application to the plant relative growth rates illustrates that the proposed calibration method produces reasonable estimates of relative growth rates and uncertainty quantification for the physical experiments.   



This work indicates several avenues for future research. First, instead of maximum likelihood estimation, Bayesian techniques can be naturally applied to the proposed method. Specifically, one could assign the priors of the calibration parameters as well as the hyperparameters in the model, and then draw samples from the posterior distribution using Markov chain Monte Carlo approaches, such as the Metropolis-Hastings sampler. 
Moreover, it is worth exploring other modeling techniques for the discrepancy function which also address the identifiability issue of the calibration parameters, such as \cite{gu2018scaled}, \cite{tuo2019adjustments}, \cite{xie2018bayesian}, and \cite{dai2018another}. These methods provide the potential to extend the proposed method with more theoretical guarantees. We are also interested in applying the proposed method to other computer models in plant biology, such as the metabolic flux models in \cite{ma2014isotopically}.
We leave it for our future work.

\vspace{5mm}
\noindent \textbf{Supplementary Materials}: The online supplementary materials contain the detailed proofs of Theorems \ref{thm:wls_consistency} and \ref{thm:asymptotic}, the detailed estimation procedure in Section \ref{sec:estimation}, supporting figures for Sections \ref{sec:illustrative} and \ref{sec:casestudy}, an \texttt{R} package \texttt{hetCalibrate} for implementing the proposed method, and the \texttt{R} code and data for reproducing the results in Sections
\ref{sec:illustrative} and \ref{sec:casestudy}.

\vspace{5mm}
\noindent \textbf{Acknowledgements}: The authors recognize funding by the Division of Chemical Sciences, Geosciences and Biosciences, Office of Basic Energy Sciences of the U.S. Department of Energy Grant DE-FG02-91ER20021 (B.W.). The authors also gratefully acknowledge helpful advice from Dr. Rui Tuo.

\bibliography{bib}

\begin{thebibliography}{}

\bibitem[Ankenman et~al., 2010]{ankenman2010stochastic}
Ankenman, B., Nelson, B.~L., and Staum, J. (2010).
\newblock Stochastic kriging for simulation metamodeling.
\newblock {\em Operations Research}, 58(2):371--382.

\bibitem[Antoniewicz et~al., 2006]{antoniewicz2006determination}
Antoniewicz, M.~R., Kelleher, J.~K., and Stephanopoulos, G. (2006).
\newblock Determination of confidence intervals of metabolic fluxes estimated
  from stable isotope measurements.
\newblock {\em Metabolic Engineering}, 8(4):324--337.

\bibitem[Bayarri et~al., 2007a]{bayarri2007computer}
Bayarri, M., Berger, J., Cafeo, J., Garcia-Donato, G., Liu, F., Palomo, J.,
  Parthasarathy, R., Paulo, R., Sacks, J., Walsh, D., et~al. (2007a).
\newblock Computer model validation with functional output.
\newblock {\em The Annals of Statistics}, 35(5):1874--1906.

\bibitem[Bayarri et~al., 2007b]{bayarri2007framework}
Bayarri, M.~J., Berger, J.~O., Paulo, R., Sacks, J., Cafeo, J.~A., Cavendish,
  J., Lin, C.-H., and Tu, J. (2007b).
\newblock A framework for validation of computer models.
\newblock {\em Technometrics}, 49(2):138--154.

\bibitem[Binois and Gramacy, 2019]{hetGP2019}
Binois, M. and Gramacy, R.~B. (2019).
\newblock {\em hetGP: Heteroskedastic Gaussian Process Modeling and Design
  under Replication}.
\newblock R package version 1.1.1.

\bibitem[Binois et~al., 2018]{binois2018practical}
Binois, M., Gramacy, R.~B., and Ludkovski, M. (2018).
\newblock Practical heteroscedastic {G}aussian process modeling for large
  simulation experiments.
\newblock {\em Journal of Computational and Graphical Statistics},
  27(4):808--821.

\bibitem[Blackman, 1919]{blackman1919compound}
Blackman, V.~H. (1919).
\newblock The compound interest law and plant growth.
\newblock {\em Annals of Botany}, 33(131):353--360.

\bibitem[Boldt et~al., 2005]{boldt2005d}
Boldt, R., Edner, C., Kolukisaoglu, {\"U}., Hagemann, M., Weckwerth, W.,
  Wienkoop, S., Morgenthal, K., and Bauwe, H. (2005).
\newblock {D-GLYCERATE 3-KINASE}, the last unknown enzyme in the
  photorespiratory cycle in arabidopsis, belongs to a novel kinase family.
\newblock {\em The Plant Cell}, 17(8):2413--2420.

\bibitem[Byrd et~al., 1995]{byrd1995limited}
Byrd, R.~H., Lu, P., Nocedal, J., and Zhu, C. (1995).
\newblock A limited memory algorithm for bound constrained optimization.
\newblock {\em SIAM Journal on Scientific Computing}, 16(5):1190--1208.

\bibitem[Caflisch, 1998]{caflisch1998monte}
Caflisch, R.~E. (1998).
\newblock {M}onte {C}arlo and quasi-{M}onte {C}arlo methods.
\newblock {\em Acta Numerica}, 7(1):1--49.

\bibitem[Dai and Chien, 2018]{dai2018another}
Dai, X. and Chien, P. (2018).
\newblock Another look at statistical calibration: a non-asymptotic theory and
  prediction-oriented optimality.
\newblock {\em arXiv preprint arXiv:1802.00021}.

\bibitem[Gneiting and Raftery, 2007]{gneiting2007strictly}
Gneiting, T. and Raftery, A.~E. (2007).
\newblock Strictly proper scoring rules, prediction, and estimation.
\newblock {\em Journal of the American Statistical Association},
  102(477):359--378.

\bibitem[Godfrey, 1978]{godfrey1978testing}
Godfrey, L.~G. (1978).
\newblock Testing for multiplicative heteroskedasticity.
\newblock {\em Journal of Econometrics}, 8(2):227--236.

\bibitem[Goldberg et~al., 1998]{goldberg1998regression}
Goldberg, P.~W., Williams, C.~K., and Bishop, C.~M. (1998).
\newblock Regression with input-dependent noise: {A} {G}aussian process
  treatment.
\newblock In {\em Advances in Neural Information Processing Systems}, pages
  493--499.

\bibitem[Gramacy et~al., 2015]{gramacy2015calibrating}
Gramacy, R.~B., Bingham, D., Holloway, J.~P., Grosskopf, M.~J., Kuranz, C.~C.,
  Rutter, E., Trantham, M., and Drake, R.~P. (2015).
\newblock Calibrating a large computer experiment simulating radiative shock
  hydrodynamics.
\newblock {\em The Annals of Applied Statistics}, 9(3):1141--1168.

\bibitem[Gu and Wang, 2018]{gu2018scaled}
Gu, M. and Wang, L. (2018).
\newblock Scaled {G}aussian stochastic process for computer model calibration
  and prediction.
\newblock {\em SIAM/ASA Journal on Uncertainty Quantification},
  6(4):1555--1583.

\bibitem[Han et~al., 2009]{han2009simultaneous}
Han, G., Santner, T.~J., and Rawlinson, J.~J. (2009).
\newblock Simultaneous determination of tuning and calibration parameters for
  computer experiments.
\newblock {\em Technometrics}, 51(4):464--474.

\bibitem[Harvey, 1976]{harvey1976estimating}
Harvey, A.~C. (1976).
\newblock Estimating regression models with multiplicative heteroscedasticity.
\newblock {\em Econometrica}, 44(3):461--465.

\bibitem[Harville, 1998]{harville1998matrix}
Harville, D.~A. (1998).
\newblock {\em Matrix Algebra from a Statistician's Perspective}.
\newblock New York: Springer-Verlag.

\bibitem[Higdon et~al., 2008]{higdon2008computer}
Higdon, D., Gattiker, J., Williams, B., and Rightley, M. (2008).
\newblock Computer model calibration using high-dimensional output.
\newblock {\em Journal of the American Statistical Association},
  103(482):570--583.

\bibitem[Higdon et~al., 2004]{higdon2004combining}
Higdon, D., Kennedy, M., Cavendish, J.~C., Cafeo, J.~A., and Ryne, R.~D.
  (2004).
\newblock Combining field data and computer simulations for calibration and
  prediction.
\newblock {\em SIAM Journal on Scientific Computing}, 26(2):448--466.

\bibitem[Hildreth and Houck, 1968]{hildreth1968some}
Hildreth, C. and Houck, J.~P. (1968).
\newblock Some estimators for a linear model with random coefficients.
\newblock {\em Journal of the American Statistical Association},
  63(322):584--595.

\bibitem[Hunt, 1982a]{hunt1982plantb}
Hunt, R. (1982a).
\newblock Plant growth analysis: second derivatives and compounded second
  derivatives of splined plant growth curves.
\newblock {\em Annals of Botany}, 50(3):317--328.

\bibitem[Hunt, 1982b]{hunt1982planta}
Hunt, R. (1982b).
\newblock {\em Plant Growth Curves. The Functional Approach to Plant Growth
  Analysis}.
\newblock Edward Arnold Ltd.

\bibitem[Kennedy and O'Hagan, 2001]{kennedy2001bayesian}
Kennedy, M.~C. and O'Hagan, A. (2001).
\newblock Bayesian calibration of computer models.
\newblock {\em Journal of the Royal Statistical Society: Series B},
  63(3):425--464.

\bibitem[Koenker and Bassett~Jr, 1982]{koenker1982robust}
Koenker, R. and Bassett~Jr, G. (1982).
\newblock Robust tests for heteroscedasticity based on regression quantiles.
\newblock {\em Econometrica}, 50(1):43--61.

\bibitem[Lee, 1992]{lee1992heteroskedasticity}
Lee, B.-J. (1992).
\newblock A heteroskedasticity test robust to conditional mean
  misspecification.
\newblock {\em Econometrica}, 60(1):159--171.

\bibitem[Loeppky et~al., 2006]{loeppky2006computer}
Loeppky, J., Bingham, D., and Welch, W. (2006).
\newblock Computer model calibration or tuning in practice.
\newblock Technical report, University of British Columbia, Vancouver, BC,
  Canada.

\bibitem[Long, 2017]{long2017note}
Long, J.~P. (2017).
\newblock A note on parameter estimation for misspecified regression models
  with heteroskedastic errors.
\newblock {\em Electronic Journal of Statistics}, 11(1):1464--1490.

\bibitem[Ma et~al., 2014]{ma2014isotopically}
Ma, F., Jazmin, L.~J., Young, J.~D., and Allen, D.~K. (2014).
\newblock Isotopically nonstationary $^{13}${C} flux analysis of changes in
  arabidopsis thaliana leaf metabolism due to high light acclimation.
\newblock {\em Proceedings of the National Academy of Sciences},
  111(47):16967--16972.

\bibitem[Mak et~al., 2018]{mak2017efficient}
Mak, S., Sung, C.-L., Wang, X., Yeh, S.-T., Chang, Y.-H., Joseph, V.~R., Yang,
  V., and Wu, C. F.~J. (2018).
\newblock An efficient surrogate model for emulation and physics extraction of
  large eddy simulations.
\newblock {\em Journal of the American Statistical Association},
  113(524):1443--1456.

\bibitem[McKay et~al., 1979]{mckay1979comparison}
McKay, M.~D., Beckman, R.~J., and Conover, W.~J. (1979).
\newblock Comparison of three methods for selecting values of input variables
  in the analysis of output from a computer code.
\newblock {\em Technometrics}, 21(2):239--245.

\bibitem[Murchie and Lawson, 2013]{murchie2013chlorophyll}
Murchie, E.~H. and Lawson, T. (2013).
\newblock Chlorophyll fluorescence analysis: a guide to good practice and
  understanding some new applications.
\newblock {\em Journal of Experimental Botany}, 64(13):3983--3998.

\bibitem[Newey and Powell, 1987]{newey1987asymmetric}
Newey, W.~K. and Powell, J.~L. (1987).
\newblock Asymmetric least squares estimation and testing.
\newblock {\em Econometrica}, 55(4):819--847.

\bibitem[Pick et~al., 2013]{pick2013}
Pick, T.~R., Bräutigam, A., Schulz, M.~A., Obata, T., Fernie, A.~R., and
  M.~Weber, A.~P. (2013).
\newblock Plgg1, a plastidic glycolate glycerate transporter, is required for
  photorespiration and defines a unique class of metabolite transporters.
\newblock {\em Proceedings of the National Acadamy of Sciences},
  110(8):3185--3190.

\bibitem[Plumlee, 2017]{plumlee2017bayesian}
Plumlee, M. (2017).
\newblock Bayesian calibration of inexact computer models.
\newblock {\em Journal of the American Statistical Association},
  112(519):1274--1285.

\bibitem[{R Core Team}, 2018]{R2018}
{R Core Team} (2018).
\newblock {\em R: A Language and Environment for Statistical Computing}.
\newblock R Foundation for Statistical Computing, Vienna, Austria.

\bibitem[Sacks et~al., 1989]{sacks1989design}
Sacks, J., Welch, W.~J., Mitchell, T.~J., and Wynn, H.~P. (1989).
\newblock Design and analysis of computer experiments.
\newblock {\em Statistical Science}, 4(4):409--423.

\bibitem[Santner et~al., 2018]{santner2003design}
Santner, T.~J., Williams, B.~J., and Notz, W.~I. (2018).
\newblock {\em The Design and Analysis of Computer Experiments (Second
  Edition)}.
\newblock Springer New York.

\bibitem[Sharkey et~al., 2007]{sharkey2007fitting}
Sharkey, T.~D., Bernacchi, C.~J., Farquhar, G.~D., and Singsaas, E.~L. (2007).
\newblock Fitting photosynthetic carbon dioxide response curves for {C}$_3$
  leaves.
\newblock {\em Plant, Cell and Environment}, 30(9):1035--1040.

\bibitem[Soetaert, 2009]{RrootSovle}
Soetaert, K. (2009).
\newblock {\em rootSolve: Nonlinear root finding, equilibrium and steady-state
  analysis of ordinary differential equations}.
\newblock R package 1.6.

\bibitem[South et~al., 2017]{south2017bile}
South, P.~F., Walker, B.~J., Cavanagh, A.~P., Rolland, V., Badger, M., and Ort,
  D.~R. (2017).
\newblock Bile acid sodium symporter bass6 can transport glycolate and is
  involved in photorespiratory metabolism in {Arabidopsis} thaliana.
\newblock {\em The Plant Cell}, 29(4):808--823.

\bibitem[Sung et~al., 2020]{sung2018calibration}
Sung, C.-L., Hung, Y., Rittase, W., Zhu, C., and Wu, C. F.~J. (2020).
\newblock Calibration for computer experiments with binary responses and
  application to cell adhesion study.
\newblock {\em Journal of the American Statistical Association}.
\newblock To appear.

\bibitem[Tuo, 2019]{tuo2019adjustments}
Tuo, R. (2019).
\newblock Adjustments to computer models via projected kernel calibration.
\newblock {\em SIAM/ASA Journal on Uncertainty Quantification}, 7(2):553--578.

\bibitem[Tuo and Wu, 2015]{tuo2015efficient}
Tuo, R. and Wu, C. F.~J. (2015).
\newblock Efficient calibration for imperfect computer models.
\newblock {\em The Annals of Statistics}, 43(6):2331--2352.

\bibitem[Tuo and Wu, 2016]{tuo2016theoretical}
Tuo, R. and Wu, C. F.~J. (2016).
\newblock A theoretical framework for calibration in computer models:
  parametrization, estimation and convergence properties.
\newblock {\em SIAM/ASA Journal on Uncertainty Quantification}, 4(1):767--795.

\bibitem[von Caemmerer, 2013]{von2013steady}
von Caemmerer, S. (2013).
\newblock Steady-state models of photosynthesis.
\newblock {\em Plant, Cell and Environment}, 36(9):1617--1630.

\bibitem[Walker et~al., 2016]{walker2016physiological}
Walker, B.~J., South, P.~F., and Ort, D.~R. (2016).
\newblock Physiological evidence for plasticity in glycolate/glycerate
  transport during photorespiration.
\newblock {\em Photosynthesis research}, 129(1):93--103.

\bibitem[Wang et~al., 2009]{wang2009bayesian}
Wang, S., Chen, W., and Tsui, K.-L. (2009).
\newblock Bayesian validation of computer models.
\newblock {\em Technometrics}, 51(4):439--451.

\bibitem[Wang and Haaland, 2019]{wang2018controlling}
Wang, W. and Haaland, B. (2019).
\newblock Controlling sources of inaccuracy in stochastic kriging.
\newblock {\em Technometrics}, 61(3):309--321.

\bibitem[Wong et~al., 2017]{wong2017frequentist}
Wong, R. K.~W., Storlie, C.~B., and Lee, T. C.~M. (2017).
\newblock A frequentist approach to computer model calibration.
\newblock {\em Journal of the Royal Statistical Society: Series B},
  79(2):635--648.

\bibitem[Xie and Xu, 2018]{xie2018bayesian}
Xie, F. and Xu, Y. (2018).
\newblock Bayesian projected calibration of computer models.
\newblock {\em arXiv preprint arXiv:1803.01231}.

\end{thebibliography}


\begin{thebibliography}{}

\bibitem[Caflisch, 1998]{suppcaflisch1988}
Caflisch, R. E. (1998).   
\newblock Monte  Carlo  and  quasi-Monte  Carlo  methods.
\newblock {\em Acta Numerica}, 7(1):1--49.



\bibitem[Harville, 1998]{suppharville1998}
Harville, D. A. (1998).
\newblock {\em Matrix Algebra from a Statistician’s Perspective}.
\newblock Springer, New York.

\bibitem[Huber, 1967]{supphuber1967}
Huber, P. J. (1967).  
\newblock The behavior of maximum likelihood estimates under nonstandard conditions. 
\newblock In {\em Proceedings of the fifth Berkeley symposium on mathematical statistics and probability}, volume 1, pages 221–233. University of California Press.

\bibitem[Mardia and Marshall, 1984]{suppmardiamarshall1984}
Mardia, K. V. and Marshall, R. J. (1984).
\newblock Maximum likelihood estimation of models forresidual covariance in spatial regression.
\newblock {\em Biometrika}, 71(1):135--146.

\bibitem[Sweeting, 1980]{suppsweeting1980}
Sweeting, T. J. (1980).
\newblock Uniform asymptotic normality of the maximum likelihood estimator.
\newblock {\em The Annals of Statistics}, 8(6):1375--1381.

\bibitem[White, 1982]{suppwhite1982}
White, H. (1982).    
\newblock Maximum  likelihood  estimation  of  misspecified  models. \newblock {\em Econometrica}, 50(1):1--25.


\end{thebibliography}

\def\spacingset#1{\renewcommand{\baselinestretch}%
{#1}\small\normalsize} \spacingset{1.3}

\newpage
\setcounter{page}{1}
\bigskip
\bigskip
\bigskip
\begin{center}
{\Large\bf Supplementary Materials for ``Calibration of Inexact Computer Models with Heteroscedastic Errors''}
\end{center}
\medskip

\setcounter{section}{0}
\setcounter{equation}{0}
\setcounter{figure}{0}
\def\theequation{S\arabic{section}.\arabic{equation}}
\def\thesection{S\arabic{section}}
\def\thefigure{S\arabic{figure}}

\section{Gradient of \eqref{eq:likelihood}}\label{append:gradient}
In this section, we derive the gradient of the log-likelihood of \eqref{eq:likelihood}, which can be used for a Newton-like optimization method. 

First, we use the Woodbury identity \citep{suppharville1998} to simplify the log-likelihood. By the Woodbury identity, it can be shown that
\begin{align*}\label{eq:woobury1}
    &\left(\mathbf{Y}_N-\mathbf{f}(\theta)\right)^T\left(\mathbf{K}_N+\boldsymbol{\Lambda}_N\right)^{-1}\left(\mathbf{Y}_N-\mathbf{f}(\theta)\right)
    =\left(\mathbf{Y}_N-\mathbf{f}(\theta)\right)^T\boldsymbol{\Lambda}_N^{-1}\left(\mathbf{Y}_N-\mathbf{f}(\theta)\right)\nonumber\\
    &-\left(\bar{\mathbf{Y}}_n-\bar{\mathbf{f}}(\theta)\right)^T\mathbf{A}_n\boldsymbol{\Lambda}_n^{-1}\left(\bar{\mathbf{Y}}_n-\bar{\mathbf{f}}(\theta)\right)+\left(\bar{\mathbf{Y}}_n-\bar{\mathbf{f}}(\theta)\right)^T\left(\mathbf{K}_n+\mathbf{A}_n^{-1}\boldsymbol{\Lambda}_n\right)^{-1}\left(\bar{\mathbf{Y}}_n-\bar{\mathbf{f}}(\theta)\right)
\end{align*}
and
\begin{equation*}\label{eq:woobury2}
    \log|\mathbf{K}_N+\boldsymbol{\Lambda}_N|=\log|\mathbf{K}_n+\mathbf{A}_n^{-1}\boldsymbol{\Lambda}_n|+\sum^n_{i=1}\left[(a_i-1)\log\lambda_i+\log a_i\right].
\end{equation*}
Then, 
\begin{align*}
    \hat{\nu}=\frac{1}{N}\left(\mathbf{Y}_N-\mathbf{f}(\theta)\right)^T\boldsymbol{\Lambda}_N^{-1}&\left(\mathbf{Y}_N-\mathbf{f}(\theta)\right)-\frac{1}{N}\left(\bar{\mathbf{Y}}_n-\bar{\mathbf{f}}(\theta)\right)^T\mathbf{A}_n\boldsymbol{\Lambda}_n^{-1}\left(\bar{\mathbf{Y}}_n-\bar{\mathbf{f}}(\theta)\right)\\&+\frac{1}{N}\left(\bar{\mathbf{Y}}_n-\bar{\mathbf{f}}(\theta)\right)^T\left(\mathbf{K}_n+\mathbf{A}_n^{-1}\boldsymbol{\Lambda}_n\right)^{-1}\left(\bar{\mathbf{Y}}_n-\bar{\mathbf{f}}(\theta)\right)
\end{align*}
and 
\begin{align*}
    \log L=-\frac{N}{2}\log\hat{\nu}-&\frac{1}{2}\log|\mathbf{K}_n+\mathbf{A}_n^{-1}\boldsymbol{\Lambda}_n|-\frac{1}{2}\sum^n_{i=1}\left[(a_i-1)\log\lambda_i+\log a_i\right]\nonumber\\
    &-\frac{n}{2}\log\hat{\nu}_{(g)}-\frac{1}{2}\log|\mathbf{K}_{(g)}+g\mathbf{A}_n^{-1}|+{\rm{Constant}}.
\end{align*}
Define $\boldsymbol{\Gamma}_n=\mathbf{K}_n+\mathbf{A}_n^{-1}\boldsymbol{\Lambda}_n$ and $\boldsymbol{\Gamma}_{(g)}=\mathbf{K}_{(g)}+g\mathbf{A}_n^{-1}$. For each component $\varphi_j$ of the lengthscale in the kernel function $k$, we have
\begin{equation*}
    \frac{\partial\log L}{\partial\varphi_j}=\frac{1}{2\hat{\nu}}\left(\bar{\mathbf{Y}}_n-\bar{\mathbf{f}}(\theta)\right)^T\boldsymbol{\Gamma}^{-1}_n\frac{\partial\mathbf{K}_n}{\partial\varphi_j}\boldsymbol{\Gamma}^{-1}_n\left(\bar{\mathbf{Y}}_n-\bar{\mathbf{f}}(\theta)\right)-\frac{1}{2}{\rm{tr}}\left(\boldsymbol{\Gamma}^{-1}_n\frac{\partial\mathbf{K}_n}{\partial\varphi_j}\right).
\end{equation*}
Here we use Monte Carlo integration \citep{suppcaflisch1988} to approximate the orthogonal kernel matrix $\mathbf{K}_n$. Suppose that the samples $\xi_1,\ldots,\xi_m$ are uniformly drawn from $\chi$, and denote that $w(x)=(k_0(x,\xi_1), \ldots,k_0(x,\xi_m))^T\in\mathbb{R}^{m\times 1}$, $\mathbf{w}=(w(\bar{x}_1),\ldots,w(\bar{x}_n)))\in\mathbb{R}^{m\times n}$, $\mathbf{F}_{\theta}=(\frac{\partial f(\xi_1,\theta)}{\partial \theta^T},\ldots, \frac{\partial f(\xi_m,\theta)}{\partial \theta^T})^T\in\mathbb{R}^{m\times q}$, $\mathbf{W}=(k_0(\xi_i,\xi_j))_{1\leq i,j\leq m}\in\mathbb{R}^{m\times m}$, and $\mathbf{K}_0=(k_0(\bar{x}_i,\bar{x}_j))_{1\leq i,j\leq n}\in\mathbb{R}^{n\times n}$. Then, by following \eqref{eq:OGP} and approximating the integration by the Monte Carlo samples, we have
\begin{equation*}
\mathbf{K}_n=\mathbf{K}_0-\mathbf{w}^T\mathbf{F}_{\theta}(\mathbf{F}^T_{\theta}\mathbf{W}\mathbf{F}_{\theta})^{-1}\mathbf{F}^T_{\theta}\mathbf{w}.
\end{equation*}
Thus, we have
\begin{align*}
    \frac{\partial\mathbf{K}_n}{\partial\varphi_j}=\frac{\partial\mathbf{K}_0}{\partial\varphi_j}-&2\frac{\partial\mathbf{w}^T}{\partial\varphi_j}\mathbf{F}_{\theta}\left(\mathbf{F}_{\theta}^T\mathbf{W}\mathbf{F}_{\theta}\right)^{-1}\mathbf{F}_{\theta}\mathbf{w}\\
    &+\mathbf{w}^T\mathbf{F}_{\theta}\left(\mathbf{F}_{\theta}^T\mathbf{W}\mathbf{F}_{\theta}\right)^{-1}\left(\mathbf{F}_{\theta}^T\frac{\partial\mathbf{W}}{\partial\varphi_j}\mathbf{F}_{\theta}\right)\left(\mathbf{F}_{\theta}^T\mathbf{W}\mathbf{F}_{\theta}\right)^{-1}\mathbf{F}_{\theta}\mathbf{w}.
\end{align*}
For common choices of kernels for $k_0$, such as Gaussian or Mat{\'e}rn kernels, the derivative, $\partial\mathbf{K}_0/\partial\varphi_j$, can be expressed in a closed form.

For the latent variance parameters, $\delta_i$, in $\boldsymbol{\Delta}_n$, we have 
\begin{equation*}
    \frac{\partial\log L}{\partial\boldsymbol{\Delta}_n}=\frac{\partial\boldsymbol{\Lambda}_n}{\partial\boldsymbol{\Delta}_n}\frac{\partial\log L}{\partial\boldsymbol{\Lambda}_n}-\frac{\boldsymbol{\Gamma}^{-1}_{(g)}\boldsymbol{\Delta}_n}{\hat{\nu}_{(g)}}=\boldsymbol{\Lambda}_n\mathbf{K}_{(g)}\boldsymbol{\Gamma}^{-1}_{(g)}\frac{\partial\log L}{\partial\boldsymbol{\Lambda}_n}-\frac{\boldsymbol{\Gamma}^{-1}_{(g)}\boldsymbol{\Delta}_n}{\hat{\nu}_{(g)}},
\end{equation*}
where 
\begin{equation*}
    \frac{\partial\log L}{\partial\boldsymbol{\Lambda}_n}=\frac{1}{2}\frac{\mathbf{A}_n\mathbf{S}\boldsymbol{\Lambda}^{-2}_n+\mathbf{A}^{-1}_n\text{diag}(\boldsymbol{\Gamma}^{-1}_n\bar{\mathbf{Y}}_n)^2}{\hat{\nu}}-\frac{\mathbf{A}_n-\mathbf{I}_n}{2}\boldsymbol{\Lambda}^{-1}_n-\frac{1}{2}\mathbf{A}^{-1}_n\text{diag}(\boldsymbol{\Gamma}^{-1}_n),
\end{equation*}
where $\mathbf{S}=\text{diag}(s_1^2,\ldots,s_n^2)$ and $s_i=\sum^{a_i}_{j=1}(y^{(j)}_{i}-\bar{y}_i)^2/a_i$.

For the each component $\phi_j$ of the lengthscale $\boldsymbol{\phi}$ in the kernel function $k_{(g)}$ of the noise process, we have
\begin{align*}
    \frac{\partial\log L}{\partial\phi_j}=&\left[\frac{\partial\mathbf{K}_{(g)}}{\partial\phi_j}-\mathbf{K}_{(g)}\boldsymbol{\Gamma}^{-1}_{(g)}\frac{\partial\mathbf{K}_{(g)}}{\partial\phi_j}\right]\boldsymbol{\Gamma}^{-1}_{(g)}\boldsymbol{\Delta}_n\boldsymbol{\Lambda}_n\times\frac{\partial\log L}{\partial\boldsymbol{\Lambda}_n}\\
    &+\frac{1}{2\hat{\nu}_{(g)}}\boldsymbol{\Delta}_n^T\boldsymbol{\Gamma}^{-1}_{(g)}\frac{\partial\mathbf{K}_{(g)}}{\partial\phi_j}\boldsymbol{\Gamma}^{-1}_{(g)}\boldsymbol{\Delta}_n-{\rm{tr}}\left(\boldsymbol{\Gamma}^{-1}_{(g)}\frac{\partial\mathbf{K}_{(g)}}{\partial\phi_j}\right).
\end{align*}
Similarly, for common choices of kernels for $k_{(g)}$, such as Gaussian or Mat{\'e}rn kernels, the derivative, $\partial\mathbf{K}_{(g)}/\partial\phi_j$, has a closed form.

For the nugget parameter $g$, we have 
\begin{align*}
    \frac{\partial\log L}{\partial g}=-\mathbf{K}_{(g)}\boldsymbol{\Gamma}^{-1}_{(g)}\mathbf{A}^{-1}_n\boldsymbol{\Gamma}^{-1}_{(g)}\boldsymbol{\Delta}_n\boldsymbol{\Lambda}_n\times\frac{\partial\log L}{\partial\boldsymbol{\Lambda}_n}+\frac{1}{2\hat{\nu}_{(g)}}\boldsymbol{\Delta}_n^T\boldsymbol{\Gamma}^{-1}_{(g)}\mathbf{A}^{-1}_n\boldsymbol{\Gamma}^{-1}_{(g)}\boldsymbol{\Delta}_n-{\rm{tr}}\left(\mathbf{A}^{-1}_n\boldsymbol{\Gamma}^{-1}_{(g)}\right)
\end{align*}

Finally, for each component $\theta_j$ of the calibration parameter $\theta$,
\begin{equation*}
    \frac{\partial\log L}{\partial \theta_j}=-\frac{1}{2\hat{\nu}}\frac{\partial N\hat{\nu}}{\partial \theta_j}-\frac{1}{2}{\rm{tr}}\left(\boldsymbol{\Gamma}^{-1}_n\frac{\partial\mathbf{K}_n}{\partial\theta_j}\right),
\end{equation*}
where
\begin{align*}
\frac{\partial N\hat{\nu}}{\partial \theta_j}=&-2\boldsymbol{\Lambda}^{-1}_N(\mathbf{Y}_N-\mathbf{f}(\theta))^T\frac{\partial\mathbf{f}(\theta)}{\partial\theta_j}+2\mathbf{A}_n\boldsymbol{\Lambda}^{-1}_n(\bar{\mathbf{Y}}_n-\bar{\mathbf{f}}(\theta))^T\frac{\partial\bar{\mathbf{f}}(\theta)}{\partial\theta_j}\\
    &-2\boldsymbol{\Gamma}^{-1}_n\left(\bar{\mathbf{Y}}_n-\bar{\mathbf{f}}(\theta)\right)^T\frac{\partial\bar{\mathbf{f}}(\theta)}{\partial\theta_j}-\left(\bar{\mathbf{Y}}_n-\bar{\mathbf{f}}(\theta)\right)^T\boldsymbol{\Gamma}^{-1}_n\frac{\partial\mathbf{K}_n}{\partial\theta_j}\boldsymbol{\Gamma}^{-1}_n\left(\bar{\mathbf{Y}}_n-\bar{\mathbf{f}}(\theta)\right),
\end{align*}
and 
\begin{align*}
\frac{\partial\mathbf{K}_n}{\partial\theta_j}=-2\mathbf{w}^T\frac{\partial\mathbf{F}}{\partial\theta_j}\left(\mathbf{F}^T\mathbf{W}\mathbf{F}\right)^{-1}\mathbf{F}\mathbf{w}+2\mathbf{w}^T\mathbf{F}\left(\mathbf{F}^T\mathbf{W}\mathbf{F}\right)^{-1}\left(\frac{\partial\mathbf{F}^T}{\partial\theta_j}\mathbf{W}\mathbf{F}\right)\left(\mathbf{F}^T\mathbf{W}\mathbf{F}\right)^{-1}\mathbf{F}\mathbf{w}.
\end{align*}

\section{Mathematical Proofs}\label{append:theorem}

\subsection{Proof of Theorme \ref{thm:wls_consistency}}\label{append:theorem_inconsistency}
Suppose that we observe $\bar{y}_1,\ldots,\bar{y}_n$ which are generated from the model $\bar{y}_i=\zeta(\bar{x}_i)+\epsilon_i$, where $\epsilon_i\sim\mathcal{N}(0,r(\bar{x}_i))$. It is easy to show that the WLS estimator $\hat{\theta}_{\rm{WLS}}$ is the maximum likelihood estimator (MLE) of $\theta$ under the misspecified model $\bar{y}_i=f(\bar{x}_i,\theta)+\epsilon_i$. Thus, under the assumption $0<r(x)<\infty$ which suffices to satisfy the regularity conditions in \cite{suppwhite1982}, the MLE converges almost surely to $\theta'$ which uniquely minimizes Kullback-Liebler divergence \citep{supphuber1967,suppwhite1982}. That is,
\begin{align*}
    \theta'=\arg\min_{\theta\in\Theta}\mathbb{E}\left[\log\frac{ \prod^n_{i=1}h_0(\bar{y}_i)}{\prod^n_{i=1}h_1(\bar{y}_i|\theta)}\right],
\end{align*}
where $h_0$ and $h_1$ are the density functions of $\bar{y}_i$ under the misspecified model and the true model, respectively. Since both models follow a normal distribution, we have 
\begin{align*}
    \mathbb{E}\left[\log\frac{\prod^n_{i=1}h_0(\bar{y}_i) }{ \prod^n_{i=1}h_1(\bar{y}_i|\theta)}\right]&=\mathbb{E}\left[\log\frac{ \exp\left(-\sum^n_{i=1}\frac{(\bar{y}_i-\zeta(\bar{x}_i))^2}{2r(\bar{x}_i)}\right)}{ \exp\left(-\sum^n_{i=1}\frac{(\bar{y}_i-f(\bar{x}_i,\theta))^2}{2r(\bar{x}_i)}\right)}\right]\\
    &=\sum^n_{i=1}\frac{1}{2r(\bar{x}_i)}\left(2(\zeta(\bar{x}_i)-f(\bar{x}_i,\theta))\mathbb{E}[\bar{y}_i]-(\zeta(\bar{x}_i)^2-f(\bar{x}_i,\theta)^2)\right)\\
    &=\sum^n_{i=1}\frac{1}{2r(\bar{x}_i)}\left(2(\zeta(\bar{x}_i)-f(\bar{x}_i,\theta))\zeta(\bar{x}_i)-(\zeta(\bar{x}_i)^2-f(\bar{x}_i,\theta)^2)\right)\\
    &=\sum^n_{i=1}\frac{\left(\zeta(\bar{x}_i)-f(\bar{x}_i,\theta)\right)^2}{2r(\bar{x}_i)}.    
\end{align*}
By the strong law of large numbers, we have 
$$\sum^n_{i=1}\frac{\left(\zeta(\bar{x}_i)-f(\bar{x}_i,\theta)\right)^2}{2r(\bar{x}_i)}\overset{a. s.}{\longrightarrow}\mathbb{E}\left[\frac{\left(\zeta(X)-f(X,\theta)\right)^2}{2r(X)}\right].$$
Therefore,
\begin{align*}
    \hat{\theta}_{\text{WLS}}\overset{a. s.}{\longrightarrow}\theta'=\arg\min_{\theta\in\Theta}\mathbb{E}\left[\frac{(\zeta(X)-f(X,\theta))^2}{r(X)}\right].
\end{align*}

\subsection{Proof of Theorem \ref{thm:asymptotic}}\label{append:theorem_asymptotic}
We first derive the information matrix $\mathbf{B}(\boldsymbol{\omega})$, and then give the regularity conditions for the theorem, and finally apply the results of \cite{suppsweeting1980} and \cite{suppmardiamarshall1984}.

For notational simplicity, we denote $\boldsymbol{\omega}=(\boldsymbol{\omega}_1, \boldsymbol{\omega}_2, \boldsymbol{\omega}_3,\boldsymbol{\omega}_4)$, where $\boldsymbol{\omega}_1=\theta, \boldsymbol{\omega}_2=(\boldsymbol{\psi},\nu),\boldsymbol{\omega}_3=(\boldsymbol{\phi},g,\nu_{(g)})$, and $\boldsymbol{\omega}_4=(\delta_1,\ldots,\delta_n)$, and their vector sizes are $m_1,m_2,m_3$ and $m_4$, respectively, with the total size $m=m_1+m_2+m_3+m_4$. We further denote $\boldsymbol{\Sigma}=\nu(\mathbf{K}_N+\boldsymbol{\Lambda}_N)$, $\mathbf{V}=\nu_{(g)}(\mathbf{K}_{(g)}+g\mathbf{A}_n^{-1})$, $\boldsymbol{\Gamma}_{(g)}=\mathbf{K}_{(g)}+g\mathbf{A}_n^{-1}$, and $\mathbf{z}=\mathbf{Y}_N-\mathbf{f}(\theta)$. Then, the log-likelihood can be rewritten as 
\begin{align*}
    \log L(\boldsymbol{\omega})=\text{constant}&-\frac{1}{2}\log|\boldsymbol{\Sigma}|-\frac{1}{2}\mathbf{z}^T\boldsymbol{\Sigma}^{-1}\mathbf{z}-\frac{1}{2}\log|\mathbf{V}|-\frac{1}{2}\boldsymbol{\Delta}^T\mathbf{V}^{-1}\boldsymbol{\Delta}.
\end{align*}
Its second derivatives can be derived as follows. For each component of $\boldsymbol{\omega}_1$,
\begin{align*}
    \frac{\partial^2 \log L(\boldsymbol{\omega})}{\partial\omega_{1i}\partial\omega_{1j}}=&-\frac{1}{2}\text{tr}\left(\boldsymbol{\Sigma}^{-1}\boldsymbol{\Sigma}^{\boldsymbol{\omega}_1}_{ij}+\boldsymbol{\Sigma}^i_{\boldsymbol{\omega}_1}\boldsymbol{\Sigma}^{\boldsymbol{\omega}_1}_{j}\right)+\mathbf{f}_{ij}^T\boldsymbol{\Sigma}^{-1}\mathbf{z}+\mathbf{f}_{j}^T\boldsymbol{\Sigma}^i_{\boldsymbol{\omega}_1}\mathbf{z}\\&-\mathbf{f}_{j}^T\boldsymbol{\Sigma}^{-1}\mathbf{f}_{i}+\mathbf{f}_{i}^T\boldsymbol{\Sigma}^j_{\boldsymbol{\omega}_1}\mathbf{z}-\frac{1}{2}\mathbf{z}^T\boldsymbol{\Sigma}^{ij}_{\boldsymbol{\omega}_1}\mathbf{z},
\end{align*}
where 
\begin{align*}
    \boldsymbol{\Sigma}^{\boldsymbol{\omega}_1}_i&=\partial \boldsymbol{\Sigma}/\partial\omega_{1i},\\
    \boldsymbol{\Sigma}_{\boldsymbol{\omega}_1}^i&=\partial \boldsymbol{\Sigma}^{-1}/\partial\omega_{1i}=-\boldsymbol{\Sigma}^{-1}\boldsymbol{\Sigma}^{\boldsymbol{\omega}_1}_i\boldsymbol{\Sigma}^{-1},\\
    \boldsymbol{\Sigma}^{\boldsymbol{\omega}_1}_{ij}&=\partial^2\boldsymbol{\Sigma}/\partial\omega_{1i}\partial\omega_{1j},\\
    \boldsymbol{\Sigma}_{\boldsymbol{\omega}_1}^{ij}&=\partial^2\boldsymbol{\Sigma}^{-1}/\partial\omega_{1i}\partial\omega_{1j}\\&=\boldsymbol{\Sigma}^{-1}(\boldsymbol{\Sigma}^{\boldsymbol{\omega}_1}_i\boldsymbol{\Sigma}^{-1}\boldsymbol{\Sigma}^{\boldsymbol{\omega}_1}_j+\boldsymbol{\Sigma}^{\boldsymbol{\omega}_1}_j\boldsymbol{\Sigma}^{-1}\boldsymbol{\Sigma}^{\boldsymbol{\omega}_1}_i-\boldsymbol{\Sigma}^{\boldsymbol{\omega}_1}_{ij})\boldsymbol{\Sigma}^{-1}.
\end{align*}
We denote $\boldsymbol{\Sigma}^{\boldsymbol{\omega}_t}_i$, $\boldsymbol{\Sigma}_{\boldsymbol{\omega}_t}^i$, $\boldsymbol{\Sigma}^{\boldsymbol{\omega}_t}_{ij}$, and $\boldsymbol{\Sigma}_{\boldsymbol{\omega}_t}^{ij}$ for $t=2,3,4$ in a similar manner.  For each component of $\boldsymbol{\omega}_1$ and $\boldsymbol{\omega}_2$,
\begin{align*}
    \frac{\partial^2 \log L(\boldsymbol{\omega})}{\partial\omega_{2i}\partial\omega_{1j}}=&-\frac{1}{2}\text{tr}\left(\boldsymbol{\Sigma}^{-1}\boldsymbol{\Sigma}^{\boldsymbol{\omega}_2\boldsymbol{\omega}_1}_{ij}+\boldsymbol{\Sigma}^i_{\boldsymbol{\omega}_2}\boldsymbol{\Sigma}^{\boldsymbol{\omega}_1}_{j}\right)+\mathbf{f}_{j}^T\boldsymbol{\Sigma}^i_{\boldsymbol{\omega}_2}\mathbf{z}-\frac{1}{2}\mathbf{z}^T\boldsymbol{\Sigma}^{ij}_{\boldsymbol{\omega}_2\boldsymbol{\omega}_1}\mathbf{z},
\end{align*}
where 
\begin{align*}
    \boldsymbol{\Sigma}^{\boldsymbol{\omega}_2\boldsymbol{\omega}_1}_{ij}&=\partial^2\boldsymbol{\Sigma}/\partial\omega_{2i}\partial\omega_{1j},\\
    \boldsymbol{\Sigma}_{\boldsymbol{\omega}_2\boldsymbol{\omega}_1}^{ij}&=\partial^2\boldsymbol{\Sigma}^{-1}/\partial\omega_{2i}\partial\omega_{1j}\\&=\boldsymbol{\Sigma}^{-1}(\boldsymbol{\Sigma}^{\boldsymbol{\omega}_2}_i\boldsymbol{\Sigma}^{-1}\boldsymbol{\Sigma}^{\boldsymbol{\omega}_1}_j+\boldsymbol{\Sigma}^{\boldsymbol{\omega}_1}_j\boldsymbol{\Sigma}^{-1}\boldsymbol{\Sigma}^{\boldsymbol{\omega}_2}_i-\boldsymbol{\Sigma}^{\boldsymbol{\omega}_2\boldsymbol{\omega}_1}_{ij})\boldsymbol{\Sigma}^{-1}.
\end{align*}
We denote $\boldsymbol{\Sigma}^{\boldsymbol{\omega}_{t_1}\boldsymbol{\omega}_{t_2}}_{ij}$ and $\boldsymbol{\Sigma}_{\boldsymbol{\omega}_{t_1}\boldsymbol{\omega}_{t_2}}^{ij}$ for any $t_1,t_2$ in a similar manner. We further denote $\mathbf{V}^{\boldsymbol{\omega}_t}_i$, $\mathbf{V}_{\boldsymbol{\omega}_t}^i$, $\mathbf{V}^{\boldsymbol{\omega}_t}_{ij}$, $\mathbf{V}_{\boldsymbol{\omega}_t}^{ij}$, $\mathbf{V}^{\boldsymbol{\omega}_{t_1}\boldsymbol{\omega}_{t_2}}_{ij}$ and $\mathbf{V}_{\boldsymbol{\omega}_{t_1}\boldsymbol{\omega}_{t_2}}^{ij}$ in a similar manner for the matrix $\mathbf{V}$. The rest of the second derived as follows,
\begin{align*}
     \frac{\partial^2 \log L(\boldsymbol{\omega})}{\partial\omega_{3i}\partial\omega_{1j}}=&-\frac{1}{2}\text{tr}\left(\boldsymbol{\Sigma}^{-1}\boldsymbol{\Sigma}^{\boldsymbol{\omega}_3\boldsymbol{\omega}_1}_{ij}+\boldsymbol{\Sigma}^i_{\boldsymbol{\omega}_3}\boldsymbol{\Sigma}^{\boldsymbol{\omega}_1}_{j}\right)+\mathbf{f}_{j}^T\boldsymbol{\Sigma}^i_{\boldsymbol{\omega}_3}\mathbf{z}-\frac{1}{2}\mathbf{z}^T\boldsymbol{\Sigma}^{ij}_{\boldsymbol{\omega}_3\boldsymbol{\omega}_1}\mathbf{z},  \\
      \frac{\partial^2 \log L(\boldsymbol{\omega})}{\partial\omega_{4i}\partial\omega_{1j}}=&-\frac{1}{2}\text{tr}\left(\boldsymbol{\Sigma}^{-1}\boldsymbol{\Sigma}^{\boldsymbol{\omega}_4\boldsymbol{\omega}_1}_{ij}+\boldsymbol{\Sigma}^i_{\boldsymbol{\omega}_4}\boldsymbol{\Sigma}^{\boldsymbol{\omega}_1}_{j}\right)+\mathbf{f}_{j}^T\boldsymbol{\Sigma}^i_{\boldsymbol{\omega}_4}\mathbf{z}-\frac{1}{2}\mathbf{z}^T\boldsymbol{\Sigma}^{ij}_{\boldsymbol{\omega}_4\boldsymbol{\omega}_1}\mathbf{z}, 
\end{align*}

\begin{align*}
     \frac{\partial^2 \log L(\boldsymbol{\omega})}{\partial\omega_{2i}\partial\omega_{2j}}=&-\frac{1}{2}\text{tr}\left(\boldsymbol{\Sigma}^{-1}\boldsymbol{\Sigma}^{\boldsymbol{\omega}_2}_{ij}+\boldsymbol{\Sigma}^i_{\boldsymbol{\omega}_2}\boldsymbol{\Sigma}^{\boldsymbol{\omega}_2}_{j}\right)-\frac{1}{2}\mathbf{z}^T\boldsymbol{\Sigma}^{ij}_{\boldsymbol{\omega}_2}\mathbf{z},  \\
      \frac{\partial^2 \log L(\boldsymbol{\omega})}{\partial\omega_{3i}\partial\omega_{2j}}=&-\frac{1}{2}\text{tr}\left(\boldsymbol{\Sigma}^{-1}\boldsymbol{\Sigma}^{\boldsymbol{\omega}_3\boldsymbol{\omega}_2}_{ij}+\boldsymbol{\Sigma}^i_{\boldsymbol{\omega}_3}\boldsymbol{\Sigma}^{\boldsymbol{\omega}_2}_{j}\right)-\frac{1}{2}\mathbf{z}^T\boldsymbol{\Sigma}^{ij}_{\boldsymbol{\omega}_3\boldsymbol{\omega}_2}\mathbf{z},\\
       \frac{\partial^2 \log L(\boldsymbol{\omega})}{\partial\omega_{4i}\partial\omega_{2j}}=&-\frac{1}{2}\text{tr}\left(\boldsymbol{\Sigma}^{-1}\boldsymbol{\Sigma}^{\boldsymbol{\omega}_4\boldsymbol{\omega}_2}_{ij}+\boldsymbol{\Sigma}^i_{\boldsymbol{\omega}_4}\boldsymbol{\Sigma}^{\boldsymbol{\omega}_2}_{j}\right)-\frac{1}{2}\mathbf{z}^T\boldsymbol{\Sigma}^{ij}_{\boldsymbol{\omega}_4\boldsymbol{\omega}_2}\mathbf{z},\\     
      \frac{\partial^2 \log L(\boldsymbol{\omega})}{\partial\omega_{3i}\partial\omega_{3j}}=&-\frac{1}{2}\text{tr}\left(\boldsymbol{\Sigma}^{-1}\boldsymbol{\Sigma}^{\boldsymbol{\omega}_3}_{ij}+\boldsymbol{\Sigma}^i_{\boldsymbol{\omega}_3}\boldsymbol{\Sigma}^{\boldsymbol{\omega}_3}_{j}\right)-\frac{1}{2}\mathbf{z}^T\boldsymbol{\Sigma}^{ij}_{\boldsymbol{\omega}_3}\mathbf{z}\\
      &-\frac{1}{2}\text{tr}\left(\mathbf{V}^{-1}\mathbf{V}^{\boldsymbol{\omega}_3}_{ij}+\mathbf{V}^i_{\boldsymbol{\omega}_3}\mathbf{V}^{\boldsymbol{\omega}_3}_{j}\right)-\frac{1}{2}\boldsymbol{\Delta}^T\mathbf{V}^{ij}_{\boldsymbol{\omega}_3}\boldsymbol{\Delta},\\
      \frac{\partial^2 \log L(\boldsymbol{\omega})}{\partial\omega_{4i}\partial\omega_{3j}}=&-\frac{1}{2}\text{tr}\left(\boldsymbol{\Sigma}^{-1}\boldsymbol{\Sigma}^{\boldsymbol{\omega}_4\boldsymbol{\omega}_3}_{ij}+\boldsymbol{\Sigma}^i_{\boldsymbol{\omega}_4}\boldsymbol{\Sigma}^{\boldsymbol{\omega}_3}_{j}\right)-\frac{1}{2}\mathbf{z}^T\boldsymbol{\Sigma}^{ij}_{\boldsymbol{\omega}_4\boldsymbol{\omega}_3}\mathbf{z}-\mathbf{e}_i^T\mathbf{V}^{j}_{\boldsymbol{\omega}_3}\boldsymbol{\Delta},\\
      \frac{\partial^2 \log L(\boldsymbol{\omega})}{\partial\omega_{4i}\partial\omega_{4j}}=&-\frac{1}{2}\text{tr}\left(\boldsymbol{\Sigma}^{-1}\boldsymbol{\Sigma}^{\boldsymbol{\omega}_4}_{ij}+\boldsymbol{\Sigma}^i_{\boldsymbol{\omega}_4}\boldsymbol{\Sigma}^{\boldsymbol{\omega}_4}_{j}\right)-\frac{1}{2}\mathbf{z}^T\boldsymbol{\Sigma}^{ij}_{\boldsymbol{\omega}_4}\mathbf{z}-\mathbf{e}_i^T\mathbf{V}^{-1}\mathbf{e}_j,\\      
\end{align*}
where $\mathbf{e}_i$ is a unit-vector where $i$-th element is one.

Thus, by the fact that $\mathbb{E}[\mathbf{z}]=\mathbf{0}$ and $\mathbb{E}[\mathbf{z}^T\mathbf{M}\mathbf{z}]=\mathbf{M}\boldsymbol{\Sigma}$ for an $N$-dimensional symmetric matrix $\mathbf{M}$, we have the information matrix 
\begin{equation*}
    \mathbf{B}_N=-\mathbb{E}\left[\frac{\partial^2\log L(\boldsymbol{\omega})}{\partial\boldsymbol{\omega}\partial\boldsymbol{\omega}^T}\right]=\mathbf{B}_0+\left[\begin{array}{cccc}
        \mathbf{B}_{11} & \mathbf{0} & \mathbf{0} & \mathbf{0}\\
         \mathbf{0} & \mathbf{0} & \mathbf{0} & \mathbf{0}\\
        \mathbf{0} & \mathbf{0} & \mathbf{B}_{33} & \mathbf{B}^T_{43}\\
         \mathbf{0}& \mathbf{0} & \mathbf{B}_{43} & \mathbf{B}_{44}
    \end{array}\right],
\end{equation*}
where $\mathbf{B}_0\in\mathbb{R}^{m\times m}, \mathbf{B}_{11}\in\mathbb{R}^{m_1\times m_1}, \mathbf{B}_{33}\in\mathbb{R}^{m_3\times m_3}, \mathbf{B}_{43}\in\mathbb{R}^{m_4\times m_3},\mathbf{B}_{44}\in\mathbb{R}^{m_4\times m_4}$ and 
\begin{align*}
    (\mathbf{B}_0)_{ij}=\frac{1}{2}&\text{tr}\left(\boldsymbol{\Sigma}^{-1}\boldsymbol{\Sigma}^{\boldsymbol{\omega}}_{i}\boldsymbol{\Sigma}^{-1}\boldsymbol{\Sigma}^{\boldsymbol{\omega}}_{j}\right),\text{ where }\boldsymbol{\Sigma}^{\boldsymbol{\omega}}_i=\partial \boldsymbol{\Sigma}/\partial\omega_{i},\\
    (\mathbf{B}_{11})_{ij}=\mathbf{f}^T_j\boldsymbol{\Sigma}^{-1}\mathbf{f}_i,&\quad (\mathbf{B}_{33})_{ij}=\frac{1}{2}\text{tr}\left(\mathbf{V}^{-1}\mathbf{V}^{\boldsymbol{\omega}_3}_{ij}+\mathbf{V}^i_{\boldsymbol{\omega}_3}\mathbf{V}^{\boldsymbol{\omega}_3}_{j}\right)+\frac{1}{2}\boldsymbol{\Delta}^T\mathbf{V}^{ij}_{\boldsymbol{\omega}_3}\boldsymbol{\Delta},\\
    (\mathbf{B}_{43})_{ij}=\mathbf{e}_i^T\mathbf{V}^{j}_{\boldsymbol{\omega}_3}\boldsymbol{\Delta},&\quad(\mathbf{B}_{44})_{ij}=\mathbf{e}_i^T\mathbf{V}^{-1}_{(g)}\mathbf{e}_j.
\end{align*} 

The derivatives, $\boldsymbol{\Sigma}^{\boldsymbol{\omega}}_i$, $\mathbf{V}_i^{\boldsymbol{\omega}_3}$, and $\mathbf{V}_{ij}^{\boldsymbol{\omega}_3}$, for each component of $\boldsymbol{\omega}$ and $\boldsymbol{\omega}_3$ are given below. We first denote $\mathbf{U}=\text{diag}(\mathbf{1}_{a_1,1},\ldots,\mathbf{1}_{a_n,1})$, where $\mathbf{1}_{k,1}$ is $k\times l$ matrix filled with ones, so we have $\mathbf{K}_N=\mathbf{U}\mathbf{K}_n\mathbf{U}^T$. Then,
\begin{align*}
    \frac{\partial \boldsymbol{\Sigma}}{\partial\theta_i}&=\nu\mathbf{U}\frac{\partial\mathbf{K}_n}{\partial\theta_i}\mathbf{U}^T,\quad \frac{\partial \boldsymbol{\Sigma}}{\partial\psi_i}=\nu\mathbf{U}\frac{\partial\mathbf{K}_n}{\partial\psi_i}\mathbf{U}^T,\quad
    \frac{\partial \boldsymbol{\Sigma}}{\partial\nu}=\mathbf{U}\frac{\partial\mathbf{K}_n}{\partial\theta_i}\mathbf{U}^T+\boldsymbol{\Lambda}_N,\\
    &\frac{\partial \boldsymbol{\Sigma}}{\partial\phi_i}=\text{diag}\left(\nu\mathbf{U}\text{diag}\left(\left[\frac{\partial\mathbf{K}_{(g)}}{\partial\phi_i}-\mathbf{K}_{(g)}\boldsymbol{\Gamma}^{-1}_{(g)}\frac{\partial\mathbf{K}_{(g)}}{\partial\phi_i}\right]\boldsymbol{\Gamma}^{-1}_{(g)}\boldsymbol{\Delta}_n\boldsymbol{\Lambda}_n\right)\mathbf{U}^T\right),\\
    &\frac{\partial \boldsymbol{\Sigma}}{\partial g}=-\text{diag}\left(\nu\mathbf{U}\text{diag}\left(\mathbf{K}_{(g)}\boldsymbol{\Gamma}^{-1}_{(g)}\mathbf{A}^{-1}_n\boldsymbol{\Gamma}^{-1}_{(g)}\boldsymbol{\Delta}_n\boldsymbol{\Lambda}_n\right)\mathbf{U}^T\right),\quad \frac{\partial \boldsymbol{\Sigma}}{\partial \nu_{(g)}}=0,\\
    &\frac{\partial \boldsymbol{\Sigma}}{\partial \delta_i}=-\text{diag}\left(\nu\mathbf{U}\text{diag}\left(\boldsymbol{\Lambda}_n\mathbf{K}_{(g)}\boldsymbol{\Gamma}^{-1}_{(g)}\mathbf{e}_i\right)\mathbf{U}^T\right),\\
    &\frac{\partial\mathbf{V}}{\partial\phi_i}=\nu_{(g)}\frac{\partial\mathbf{K}_{(g)}}{\partial\phi_i},\quad\frac{\partial\mathbf{V}}{\partial g}=\nu_{(g)}\mathbf{A}^{-1}_n,\quad\frac{\partial\mathbf{V}}{\partial \nu_{(g)}}=\boldsymbol{\Gamma}_{(g)},\quad\frac{\partial\mathbf{V}}{\partial\delta_i}=\mathbf{0},\\
    &\frac{\partial^2\mathbf{V}}{\partial \phi_i\partial\phi_i}=\nu_{(g)}\frac{\partial^2\mathbf{K}_{(g)}}{\partial\phi_i\partial\phi_j},\quad\frac{\partial^2\mathbf{V}}{\partial \nu_{(g)}\partial\phi_j}=\frac{\partial\mathbf{K}_{(g)}}{\partial\phi_j},\quad\frac{\partial^2\mathbf{V}}{\partial \nu_{(g)}\partial g}=\mathbf{A}_n^{-1},\\
    &\frac{\partial^2\mathbf{V}}{\partial g\partial\phi_j}=\frac{\partial^2\mathbf{V}}{\partial g\partial g}=\frac{\partial^2\mathbf{V}}{\partial \nu_{(g)}\partial \nu_{(g)}}=\frac{\partial^2\mathbf{V}}{\partial \delta_i\partial\phi_j}=\frac{\partial^2\mathbf{V}}{\partial \delta_i\partial g}=\frac{\partial^2\mathbf{V}}{\partial \delta_i\partial \nu_{(g)}}=\frac{\partial^2\mathbf{V}}{\partial \delta_i\partial \delta_j}=\mathbf{0},
\end{align*}
where $\partial\mathbf{K}_{g}/\partial\phi_i\partial\phi_j$ can be expressed in a closed form for common choices of kernels for $k_{g}$, such as Gaussian or Mat{\'e}rn kernels.

The regularity conditions for the asymptotic result of Theorem \ref{thm:asymptotic} are provided below.
\begin{enumerate}
    \item The kernels $k$ and $k_{(g)}$ are twice differentiable on the parameter spaces of $\theta,\boldsymbol{\varphi}$ and $\boldsymbol{\phi}$ with continuous second derivatives. 
    \item The smallest latent root of $\mathbf{B}_N$ tends to $\infty$ as $N\rightarrow\infty$.
    \item $\mathbf{B}^{-1/2}_N\left(\frac{\partial^2 \log L}{\partial\boldsymbol{\omega}\partial\boldsymbol{\omega}^T}\right)\mathbf{B}^{-1/2}_N$ converges in probability to a unit matrix. 
\end{enumerate}
Similar to \cite{suppmardiamarshall1984} which uses the general result of MLE in \cite{suppsweeting1980} to show the consistency and asymptotic normality of MLE of a Gaussian process model, we have that, under the regularity conditions, the general result of \cite{suppsweeting1980} gives that the MLE $\hat{\boldsymbol{\omega}}_N$ is consistent and asymptotically normally distributed as $N$ is sufficiently large.

\section{Supporting Figures in Sections \ref{sec:illustrative} and \ref{sec:casestudy}}\label{append:casestudyfigure}
The figures that present the calibration results in Sections \ref{sec:illustrative} and \ref{sec:casestudy} are provided in this section.

\begin{figure}[t!]
    \centering
    \includegraphics[width=\textwidth]{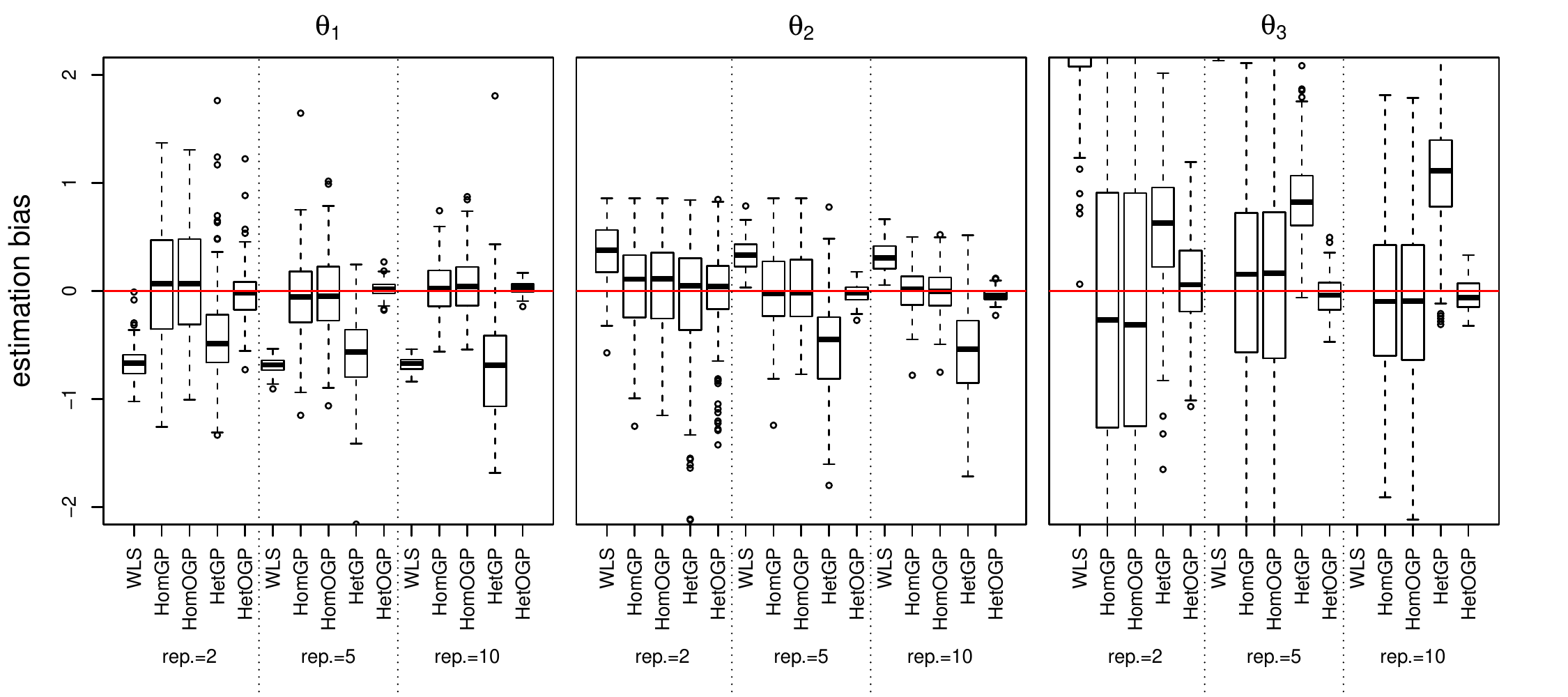}
    \caption{The estimation bias of (left) $\theta_1$; (middle) $\theta_2$; (right) $\theta_3$, with the red horizontal  line indicating zero bias. Results with 2, 5 and 10 replicates at each input location are arranged in three groups of five along the $x$-axis in each panel.}
    \label{fig:example3comparison}
\end{figure}

\begin{figure}[h]
    \centering
    \includegraphics[width=0.9\textwidth]{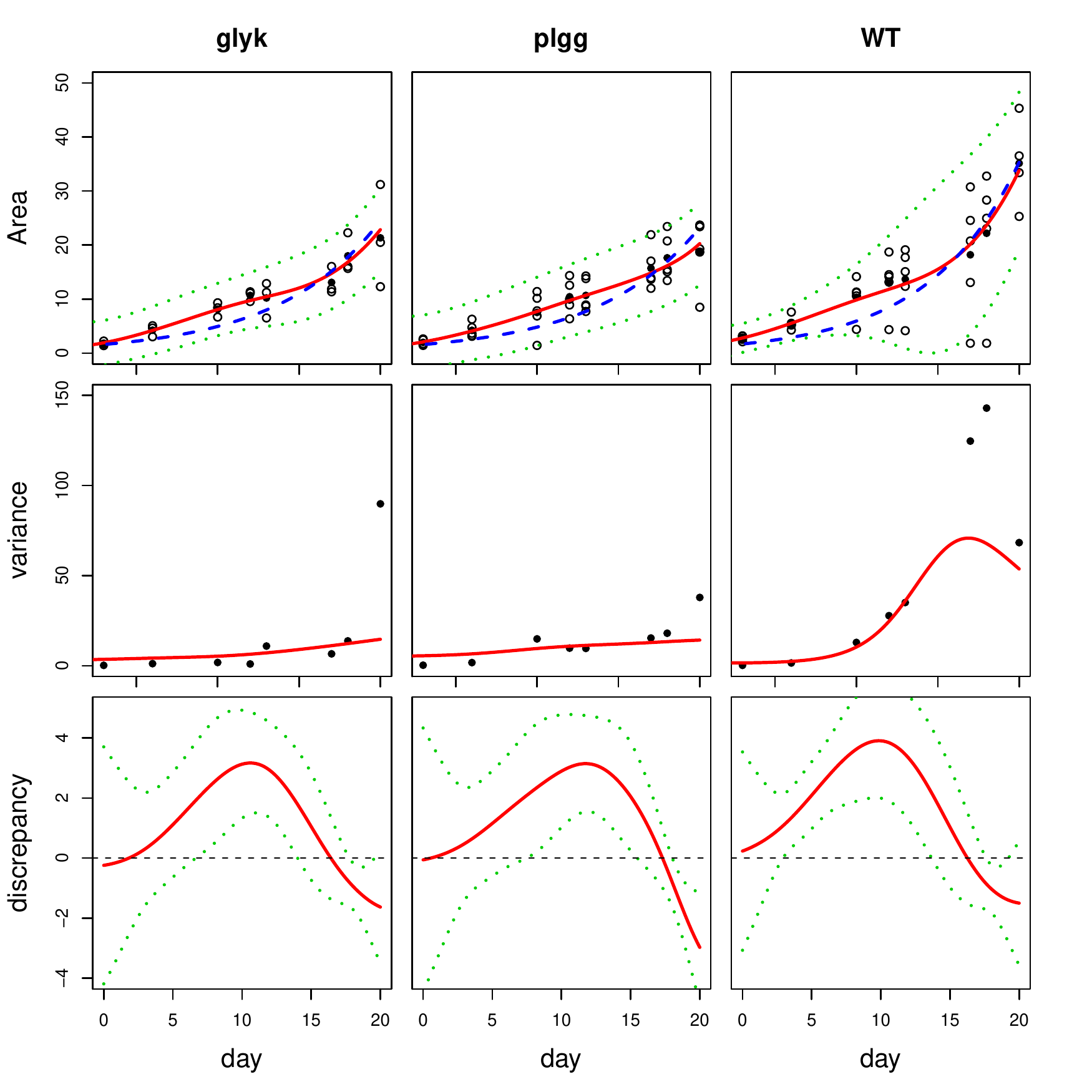}
    \caption{Calibration results of the three plant groups under ambient CO$_2$: \textit{glyk} (left), \textit{plgg} (middle), and WY (right). Top panels represent the replicates as open circles with the averaged observation $\hat{y}_i$ in filled circles at each input location, the curve $f(x,\hat{\theta})$ as a blue dashed line, and the prediction mean curve as a red solid line, with 95\% prediction intervals in green dotted lines. Middle panels represent the sample variance $\hat{r}(\bar{x}_i)$ as black points, and the fitted variance process as a red solid line. Lower panels represent the mean curve of the discrepancy function, with 95\% pointwise confidence intervals in green dotted lines.}
    \label{fig:realcalibration_ambient}
\end{figure}

\begin{figure}[h]
    \centering
    \includegraphics[width=0.9\textwidth]{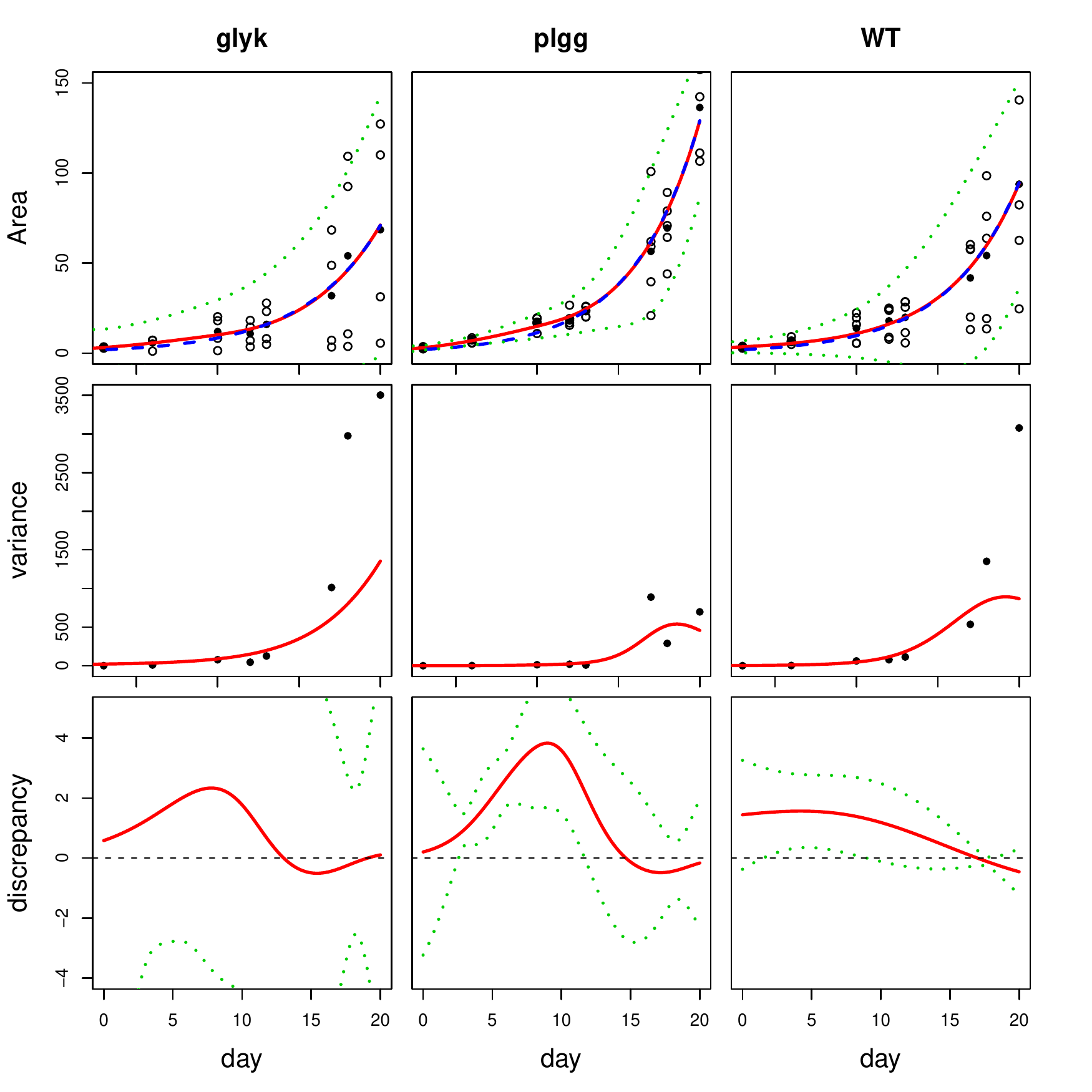}
    \caption{Calibration results of the three plant groups under high CO$_2$: \textit{glyk} (left), \textit{plgg} (middle), and WY (right). Top panels represent the replicates as open circles with the averaged observation $\hat{y}_i$ in filled circles at each input location, the curve $f(x,\hat{\theta})$ as a blue dashed line, and the prediction mean curve as a red solid line, with 95\% prediction intervals in green dotted lines. Middle panels represent the sample variance $\hat{r}(\bar{x}_i)$ as black points, and the fitted variance process as a red solid line. Lower panels represent the mean curve of the discrepancy function, with 95\% pointwise confidence intervals in green dotted lines.}
    \label{fig:realcalibration_hi}
\end{figure}

\end{document}